\def\rfr#1{eq. (\ref{#1})}

\def\derp#1#2{\rp{\partial{#1}}{\partial{#2}}}
\def\dert#1#2{\frac{{{d}}{#1}}{{{d}}{#2}}}              % derivate parziali e totali prima e seconda

\def\virg#1{``#1''}

\def\eqi{\begin{equation}}
\def\eqf{\end{equation}}
\def\eqia{\begin{eqnarray}}
\def\eqfa{\end{eqnarray}}
\def\rp#1#2{{#1\over#2}} \def\lb#1{\label{#1}}

\def\bds#1{\boldsymbol{#1}}

\documentclass[usenatbib]{mn2e}
\usepackage{amsmath}
\usepackage{amscd}
\usepackage{amssymb}
\usepackage{latexsym}
\usepackage{graphicx}
\usepackage{epsfig}

\title[Long-term time variations in exoplanets]{Classical and relativistic long-term time variations of some observables for transiting exoplanets}
\author[L. Iorio]{
L. Iorio$^{1}$\thanks{E-mail:
lorenzo.iorio@libero.it}\thanks{Present address: Viale Unit$\grave{\rm a}$ di Italia 68, 70125, Bari (BA), Italy.}\\
$^{1}$Ministero dell'Istruzione, dell'Universit\`{a} e della Ricerca (MIUR), Fellow of the Royal Astronomical Society (FRAS), Italy 
}

\begin{document}

\date{Accepted 2010 September 8. Received 2010 September 8; in original form 2010 July 16}

%\pagerange{\pageref{firstpage}--\pageref{lastpage}} \pubyear{2002}

\maketitle

\label{firstpage}

\begin{abstract}
We analytically work out the long-term, i.e. averaged over one orbital revolution, time variations $\left\langle\dot Y\right\rangle$
of some direct observable quantities $Y$
 induced by classical and general relativistic dynamical perturbations of the two-body pointlike Newtonian acceleration in the case of  transiting exoplanets moving along elliptic orbits. More specifically, the observables $Y$ with which we deal  are the transit duration $\Delta t_d$, the radial velocity $V_{\rho}$, the time interval $\Delta t_{\rm ecl}$ between primary and secondary eclipses, \textrm{and the time interval $P_{\rm tr}$ between successive primary transits}. The dynamical effects considered are the centrifugal oblateness of both the star and the planet, their tidal bulges mutually raised on each other, a distant third body X, and general relativity (both Schwarzschild and Lense-Thirring). We take into account the effects due to the  perturbations of all the Keplerian orbital elements involved in a consistent and uniform way. First, we explicitly compute their instantaneous time variations  due to the dynamical effects considered and substitute them in the general expression for the instantaneous change of $Y$; then, we take the overall average over one orbital revolution of the so-obtained instantaneous rate $\dot Y(t)$ specialized to the perturbations considered.
  \textrm{In contrast, previous published works have often employed somewhat \virg{hybrid} expressions, in which} the secular precession of, typically, the periastron only is straightforwardly inserted into instantaneous formulas.
 The transit duration is affected neither by the general relativistic Schwarzschild-type  nor by the classical tidal effects, while the bodies'centrifugal oblatenesses, a distant third body X and the general relativistic Lense-Thirring-type perturbations induce non-vanishing \textrm{long-term, harmonic} effects on $\Delta t_{d}$ also for circular orbits. For exact edge-on configurations they vanish. Both $V_{\rho}$ and $\Delta t_{\rm ecl}$ experience non-vanishing \textrm{long-term, harmonic} variations, caused by all the perturbations considered, only for non-circular orbits. \textrm{Also $P_{\rm tr}$ is affected by all of them with long-term signatures, but they do not vanish for circular orbits.}
 Numerical evaluations of the obtained results are given for a typical star-planet scenario and compared with the expected observational accuracies over a time span $\tau=10$ yr long. \textrm{Also graphical investigations of the dependence of the effects considered on the semi-major axis and the eccentricity are presented.}
  Our results  are, in principle, valid also for other astronomical scenarios. They may allow, e.g., for designing various tests of gravitational theories with natural and artificial bodies in our solar system.
\end{abstract}

%\keywords{Experimental studies of gravity; Experimental tests of gravitational theories; Extrasolar planetary systems; Main-sequence: intermediate-type stars (A and F); Stellar rotation }
%PACS: 04.80.-y; 04.80.Cc; 97.82.-j; 97.20.Ge; 97.10.Kc

\section{Introduction}\lb{intro}
%The light curves of transiting exoplanets, in conjunction with spectroscopic analysis of their radial velocities, reveal a great deal of information con

In several papers  somewhat \virg{hybrid} expressions for certain observable quantities  pertaining \textrm{to} exoplanets \textrm{are} present\textrm{ed}. In them, formulas for the long-term, i.e. averaged over one orbital revolution, precession of the periastron $\omega$ due to some dynamical perturbations are straightforwardly \textrm{substituted} in expressions which, instead, contain instantaneous values of the true anomaly $f$ of the planet. Moreover, a certain lack of uniformity in the various treatments tend to make \textrm{them}, perhaps, difficult to follow.

In this paper we propose a more uniform approach by systematically working out, in an analytic and explicit way whenever possible and/or convenient, the expressions for the long-term variations of various observables induced by a number of dynamical perturbations, both of classical and general relativistic origin. In particular, we will focus on the long-term variations of the transit duration $\Delta t_d$, the radial velocity $V_{\rho}$, the time interval $\Delta t_{\rm ecl}$ between primary and secondary eclipses, \textrm{and the time $P_{\rm tr}$ elapsed between consecutive primary transits}.
 Concerning general relativity, the issue of the measurability of certain effects of it in some extrasolar planets has been treated in \citet{Miralda,Adamsa,Adamsb,Adamsc,Iorio,Heyl,Jordan,Pal,Ragozzine,Li,Iorio010}. In particular, the perspectives in measuring the  correction to the third Kepler law have been considered by \citet{Iorio} and \citet{Ragozzine}. \citet{Miralda}, \citet{Heyl}, \citet{Jordan}, \citet{Pal} and \citet{Ragozzine} dealt with the possibility of detecting the gravitoelectric periastron precessions, while \citet{Li} looked at the gravitoelectric secular change of the mean anomaly $\mathcal{M}$ connected to the variation of the periastron time transit $t_p$. \citet{Adamsa,Adamsb,Adamsc} studied the impact of the general relativistic gravitoelectric terms on the long-term, secular interactions among multiple planetary systems. \citet{Iorio010} looked at the relativistic effects induced by the rotation of the hosting star.

In our calculation, we consistently take into account  the perturbations of all the Keplerian orbital elements in the following way.
Let us assume that
an explicit expression is available for a given observable $Y$ in such a way that it
is function of all or some Keplerian orbital elements, i.e. $Y=Y(f,\{\kappa\})$, where $\kappa$ denotes the ensemble of the Keplerian orbital elements explicitly entering $Y$ apart from the mean anomaly $\mathcal{M}$. Then, we straightforwardly compute its long-term variation as the sum of two parts. The first one is purely Keplerian, and it vanishes over one orbital period $P_{\rm b}$. The second one is due to the non-Keplerian variations of all the  orbital elements induced by the dynamical perturbation considered. The total result is, thus\footnote{\textrm{The analytical calculations have been performed with the aid of the software MATHEMATICA.}},
\eqi
\begin{array}{lll}
\left\langle\dert{Y}{t}\right\rangle & = & \rp{\Delta Y}{P_{\rm b}}=\left(\rp{1}{P_{\rm b}}\right)\int_0^{2\pi}\left[\derp{Y}{f}\dert{f}{\mathcal{M}}\dert{\mathcal{M}}{t}+\right.\\ \\
&+&\left.\sum_{\kappa}\derp{Y}{\kappa}\dert{\kappa}{t}\right]\left(\dert{t}{f}\right) df.\lb{ypa}
\end{array}
\eqf
In it,  ${d\mathcal{M}}/{dt}$ and $d\kappa/dt$ are the instantaneous variations\footnote{Actually, $d\mathcal{M}/dt$ is the sum of the Keplerian mean motion $n$ and a non-Keplerian term, as we will see later. Its Keplerian part  yields from \rfr{ypa} the Keplerian variation of $Y$.} of the Keplerian orbital elements computed with, e.g., the Gauss variation equations and evaluated onto the unperturbed Keplerian ellipse, while ${df}/{d\mathcal{M}}$ and $dt/df$ are the usual Keplerian expressions for such derivatives: see \rfr{yta} and \rfr{yga} below.
\textrm{To make easier for the reader to go through the details of the following calculations, we list in Table \ref{definizioni} all the symbols used along with their definitions. More details are also given in the text.
\begin{table*}
\caption{Symbols used in the text along with their definitions.}\label{definizioni}
\begin{tabular}{@{}lll}
\hline
Symbol &  Definition  \\
\hline
$G$ & Newtonian constant of gravitation\\
$c$ & Speed of light in vacuum \\
%
%$t$ & Time \\
%
$M$ & Mass of the primary in the two-body problem\\
$R_e$ & Equatorial radius of the primary in the two-body problem\\
$J_2$ & Adimensional quadrupole mass moment of the primary in the two-body problem\\
$S$ & Angular momentum of the primary in the two-body problem\\
$M_{\star}$  & Mass of the  star\\
$R_{\star}$  & Radius of the  star \\
$S_{\star}$  & Angular momentum of the  star\\
$J^{\star}_2$ & Adimensional quadrupole mass moment of the  star \\
$k_{2\star}$ & Love number of the  star\\
$m_{\rm p}$  & Mass of the planet\\
$r_{\rm p}$  & Radius of the planet\\
$k_{2{\rm p}}$ & Love number of the planet\\
$j_2^{\rm p}$ & Adimensional quadrupole mass moment of the planet\\
$\mathcal{R}_g\doteq G(M_{\star}+m_{\rm p})/c^2$ & Gravitoelectric characteristic length of the system\\
$\chi_g\doteq 2 G S_{\star}/c^2$ & Gravitomagnetic characteristic volume per unit time of the system\\
$U_{\rm X}$ & Tidal gravitational potential of a distant  third body X\\
$m_{\rm X}$ &  Mass of the third body X\\
$r_{\rm X}$ & Distance of the third body X from the star (assumed constant over one orbital period of the planet p)\\
$\mathcal{K}_{\rm X}\doteq Gm_{\rm X}/r^3_{\rm X}$ & Tidal parameter of the third body\\
$\bds{\hat{l}}_{\rm X}=\{l_x,l_y,l_z\}$ & Unit vector pointing towards the third body (assumed constant over one orbital period of the planet p)\\
%
%$x,y,z$ & Cartesian coordinates of the planet\\
%
$r$ & Star-planet relative distance\\
$a$ &  Semi-major axis of the orbit\\
$n\doteq \sqrt{G(M_{\star}+m_{\rm p})/a^3}$  & Keplerian mean motion of  the orbit \\
$P_{\rm b}\doteq 2\pi/n$ & Orbital period\\
$e$ &  Eccentricity of the orbit \\
$p\doteq a(1-e^2)$ & Semilatus rectum of the orbit\\
$v$ & Speed of the planet along its orbit (assumed elliptical)\\
$i$  & Inclination of the orbit  to the plane of the sky \\
$I_{\star}$ & Inclination of the angular momentum of the star to the plane of the sky\\
$\Psi_{\star}$  & Inclination of the orbit  to the equatorial plane of the star\\
$\psi_{\rm p}$   & Inclination of the orbit  to the equatorial plane of the planet\\
$f$ & True anomaly of the orbit \\
$\omega$ & Argument of pericenter of the orbit\\
$\Omega$ & Longitude of the ascending node of the orbit \\
$u\doteq \omega + f$ & Argument of latitude of the orbit \\
$\mathcal{M}$ & Mean anomaly of the orbit \\
$\{\kappa\}$ & Set of Keplerian orbital elements (apart from the mean anomaly)\\
$\eta_g\doteq \chi _g n a^{-2} (1-e^2)^{-7/2}(1+e\cos f)^3 $ & Gravitomagnetic characteristic acceleration parameter of the system\\
$\bds{\hat{R}}$ & Unit vector along the radial direction of the co-moving frame\\
$\bds{\hat{T}}$ & Unit vector along the transverse direction of the co-moving frame\\
$\bds{\hat{N}}$ & Unit vector along the normal direction of the co-moving frame\\
$\bds A$ & Perturbing acceleration caused by relativity, oblateness, etc.\\
$A_R\doteq \bds A\bds\cdot \bds{\hat{R}}$ & Radial component of the perturbing acceleration\\
$A_T\doteq \bds A\bds\cdot \bds{\hat{T}}$ & Transverse component of the perturbing acceleration\\
$A_N\doteq \bds A\bds\cdot \bds{\hat{N}}$ & Normal component of the perturbing acceleration\\
$\bds{\hat{S}}_{\star}$ & Unit vector along the angular momentum of the star\\
$\bds{\hat{L}}$ & Unit vector along the orbital angular momentum\\
$\bds{\hat{\rho}}$ & Unit vector along the line of sight (pointing towards the observer)\\
$Y$ & Generic observable quantity\\
$\dot Y$ & Instantaneous time variation of the generic observable quantity $Y$\\
$\Delta Y$ & Net change of the generic observable quantity $Y$ after one orbital revolution\\
$\left\langle \dot Y\right\rangle $ & Averaged time variation of the generic observable quantity $Y$\\
$\tau$ & Observational time interval\\
$\sigma_{\dot Y}$ & Accuracy in measuring $\dot Y$\\
$\delta\ (\overline{\delta})$ & Latitude of the transit for elliptic orbits (circular orbits)\\
%
%$\overline{\delta}$ & Latitude of the transit (circular orbits) \\
%
$\Delta t_d$ & Transit duration\\
$V_{\rho}$ & Radial velocity\\
$K$ & Semiamplitude of the radial velocity curve\\
$V_0$ & Radial velocity of the center of mass of the system\\
$\Delta t_{\rm ecl}$ &  Time span between the primary and the
secondary eclipses\\
$P_{\rm tr}$ & Time span between two consecutive primary eclipses\\
\hline
\end{tabular}
\end{table*}
}

In order to give \textrm{just some preliminary} numerical estimates of the effects computed, we will consider a rather typical star-planet scenario summarized in Table \ref{parametri}.
\begin{table}
\caption{Reference stellar and planetary parameters adopted in the text. We use \textrm{$M_{\star}=M_{\odot}$}, $R_{\star}=R_{\odot}$, $J_2^{\star}=J_2^{\odot}$ \citep{J2Sun}, $S_{\star}=S_{\odot}$\citep{Pijp1,Pijp2}, $k_{2\star}=0.03$ \citep{Claret}, $m_{\rm p}=m_{\rm Jup}$, $r_{\rm p}=r_{\rm Jup}$, $k_{2{\rm p}}=0.6$ \citep{Ragozzine}. Concerning $j_2^{\rm p}$, we computed it from $j_2^{\rm p}=(k_{2{\rm p}}/3)(n^2 r_{\rm p}^3/Gm_{\rm p})$ \citep{Ragozzine}, where we assumed that the rotational frequency of the planet is equal to the orbital one $n$ because of tidal effects. For simplicity, we assume that the equators of both the star and the planet coincide with the orbital plane, so that $\Psi_{\star}=\psi_{\rm p}=0$. Moreover, we assume an edge-on orbital configuration, i.e. $i=90$ deg.}\label{parametri}
\begin{tabular}{@{}lll}
\hline
Parameter (units) &  Numerical value  \\
\hline
$M_{\star}$ (kg) & $1.98895\times 10^{30}$\\
$R_{\star}$ (m) & $6.9599\times 10^8$ \\
$S_{\star}$ (kg m$^2$ s$^{-1}$) & $190.0\times 10^{39}$\\
$J^{\star}_2$ & $2\times 10^{-7}$ \\
$k_{2\star}$ & 0.03\\
$m_{\rm p}$ (kg) & $1.89\times 10^{27}$\\
$r_{\rm p}$ (m) & $6.991\times 10^7$\\
$k_{2{\rm p}}$ & 0.6\\
$j_2^{\rm p}$ & $3\times 10^{-4}$\\
$a$ (au)& $0.04$ \\
$n$ (s$^{-1}$) & $2.5\times 10^{-5}$\\
$P_{\rm b}$ (d) & $2.9$\\
$e$ & $0.07$ \\
$i$ (deg) & 90 \\
$\Psi_{\star}$ (deg) & 0\\
$\psi_{\rm p}$  (deg) & 0\\
\hline
\end{tabular}
\end{table}
It includes a Sun-like host star harboring a close Jupiter-sized planet moving in the equatorial plane of the star along a moderately elliptic orbit. The orbital angular momentum is  perpendicular to the line of sight, so that the orbit is in an edge-on configuration. We assume that the rotation of the planet has been locked by tidal effects to the orbital frequency $n$, and that all the spins of the system are aligned. The meaning of the symbols used in Table \ref{parametri} \textrm{has been} explained  \textrm{in Table \ref{definizioni}}. However, it must be noted that, if \textrm{on} the one hand, such a scenario is largely representative of most of the transiting exoplanets discovered so far \citep{Torres}, on the other hand some planets  showing high inclinations of their orbits to the stellar equatorial planes have been discovered in recent times. They are WASP-33b \citep{Wasp33}, WASP-2b \citep{wasp2}, WASP-5b \citep{wasp5,wasp5b}, WASP-8b \citep{wasp8}, WASP-15b \citep{wasp15}, WASP-17b \citep{Ross} and HAT-P-7b \citep{hat1,hat2}. Our general expressions are valid also for such class of objects.
\textrm{Moreover, the Kepler mission is expected to detect numerous planets with a wide distribution of both masses and semi-major axes; to a certain extent, it has already started to do that \citep{Bou}. A quantitative extension of our results to such specific exoplanetary scenarios and to others exhibiting not yet explained features which may be accommodated by the effects investigated here may be the subject of a further, more applicative paper.}

Let us note that the validity of the formulas obtained is not limited only to exoplanetary scenarios. Indeed, they may be applied also for, e.g., designing new observational tests of general relativity in our solar system with natural and artificial bodies. In principle, they can be adapted  to modified models of gravity as well. The latter aspect will be developed in further studies.

The paper is organized as follows. In Section \ref{accelez} we illustrate the main features of the classical and relativistic perturbing accelerations considered useful for our purposes. The time variations of the duration transit are worked out in Section \ref{durazza}. Section \ref{velazza} deals with the time changes of the radial velocity. Section \ref{eclizza} is devoted to the time interval elapsed between primary and secondary eclipses and its long-term temporal variations. \textrm{The time span between consecutive primary transits is investigated in Section \ref{titivu}.} In Section \ref{concluzza} we summarize our results.

\section{The perturbing accelerations}\lb{accelez}
Here we deal with a generic perturbing acceleration $\bds A$ induced by some  classical and general relativistic dynamical effects.

First, $\bds A$ has to be projected onto the radial, transverse and normal orthogonal unit vectors $\bds{\hat{R}},\bds{\hat{T}},\bds{\hat{N}}$ of the co-moving frame of the test particle orbiting the central body acting as source of the gravitational field.
Their components, in cartesian coordinates, are  \citep{Monte}
\eqi \bds{\hat{R}} =\left(
       \begin{array}{c}
          \cos\Omega\cos u\ -\cos \Psi\sin\Omega\sin u\\
          \sin\Omega\cos u + \cos \Psi\cos\Omega\sin u\\
         \sin \Psi\sin u \\
       \end{array}
     \right)\lb{ierre}
\eqf
 \eqi \bds{\hat{T}} =\left(
       \begin{array}{c}
         -\sin u\cos\Omega-\cos \Psi\sin\Omega\cos u \\
         -\sin\Omega\sin u+\cos \Psi\cos\Omega\cos u \\
         \sin \Psi\cos u \\
       \end{array}
     \right)\lb{itrav}
\eqf
\eqi \bds{\hat{N}} =\left(
       \begin{array}{c}
          \sin \Psi\sin\Omega \\
         -\sin \Psi\cos\Omega \\
         \cos \Psi\\
       \end{array}
     \right)\lb{inorm}.
\eqf
In \rfr{ierre}-\rfr{inorm}, $\Omega,\omega,\Psi$ are the longitude of the ascending node\footnote{See also \rfr{megas} below.}, the argument of pericenter, reckoned from the line of the nodes\footnote{It is the intersection of the orbital plane with the equatorial plane. See \rfr{trigga} and \rfr{megas} below.}, and the inclination of the orbital plane to the reference $\{xy\}$ plane, respectively. Moreover, $u\doteq f+\omega$ is the argument of latitude.
Subsequently, the projected components of $\bds A$ have to be evaluated onto the Keplerian ellipse
\eqi r=\rp{p}{1+e\cos f},\ p\doteq a(1-e^2),\lb{rkep}\eqf where $p$ is the semilatus rectum and $a,e$ are the semi-major axis and the eccentricity, respectively.
The cartesian coordinates of the Keplerian motion in space are
 \citep{Monte}
 \begin{equation}
{\begin{array}{lll}
 x &=& r\left(\cos\Omega\cos u\ -\cos \Psi\sin\Omega\sin u\right),\\  \\
 y &=& r\left(\sin\Omega\cos u + \cos \Psi\cos\Omega\sin u\right),\\  \\
 z &=& r\sin \Psi\sin u.
\end{array}}\lb{xyz}
 \end{equation}
Then,  $A_R,A_T,A_N$ are to be \textrm{substituted} into the right-hand-sides of the  Gauss equations for the variations of the Keplerian orbital elements. They are \citep{Roy,Soffel}
\begin{equation}
{\begin{array}{lll}
\dert{a}{t}&=& \rp{2}{n\sqrt{1-e^2}}\left[A_R e \sin f+ A_T\left(\rp{p}{r}\right)\right],\\ \\
\dert{e}{t}&=& \rp{\sqrt{1-e^2}}{na}\left\{ A_R\sin f +A_T\left[\cos f+\rp{1}{e}\left(1-\rp{r}{a}\right)\right]\right\},\\ \\
\dert{\Psi}{t}&=&\rp{1}{na\sqrt{1-e^2}}A_N\left(\rp{r}{a}\right)\cos u,\\\\
\dert{\Omega}{t}&=& \rp{1}{na\sqrt{1-e^2}\sin\Psi}A_N\left(\rp{r}{a}\right)\sin u,\\\\
\dert{\omega}{t}&=& -\cos\Psi\left(\dert{\Omega}{t}\right)+\rp{\sqrt{1-e^2}}{nae}\left[-A_R\cos f+\right.\\ \\
&+&\left.A_T\left(1+\rp{r}{p}\right)\sin f\right], \\ \\
\dert{\mathcal{M}}{t}&=& n-\rp{2}{na}A_R\left(\rp{r}{a}\right)-\rp{(1-e^2)}{nae}\left[-A_R\cos f+\right.\\ \\
&+&\left. A_T\left(1+\rp{r}{p}\right)\sin f\right],
\end{array}}\lb{Gauss}
 \end{equation}
 where  $n\doteq \sqrt{G(M_{\star}+m_{\rm p})/a^3}$ is the Keplerian mean motion  related to the orbital period by $n=2\pi/P_{\rm b}$.

 As explained in the Introduction, the right-hand-sides of \rfr{Gauss}, computed for the perturbing accelerations of the dynamical effect considered, have to be inserted into the analytic expression of the time variation $dY/dt$ of the observable $Y$  of interest which, then, must be averaged over one orbital revolution according to \rfr{ypa} by means of \citep{Roy}
\eqi df =\left(\rp{a}{r}\right)^2\sqrt{1-e^2}d\mathcal{M},\lb{yta}\eqf
and
 \eqi dt = \rp{(1-e^2)^{3/2}}{n(1+e\cos f)^2}df.\lb{yga}\eqf
\subsection{The effect of the stellar oblateness}
The external gravitational field of a rotating body undergoes departures from spherical symmetry because of the distortion of its shape due to the resulting centrifugal force.
An oblate body of equatorial radius $R_e$ and adimensional quadrupole mass moment $J_2$ affects the orbital motion of a test particle with a  non-central perturbing acceleration \citep{Cunn,Vrbik}
\eqi \bds{A}^{(J_2)}=-\rp{3J_2 R_e^2 GM}{2r^4}\left\{\left[1-5(\bds{\hat{r}}\bds\cdot\bds{\hat{S}})^2\right]\bds{\hat{r}}+2(\bds{\hat{r}}\bds\cdot\bds{\hat{S}})\bds{\hat{S}}\right\},\lb{sgorbia}\eqf
where $\hat{\bds{S}}$ is the unit vector of the body's angular momentum, chosen here as $z$ axis so that the equatorial plane is the reference $\{xy\}$. We will adopt such a choice for the coordinate axes also for the other dynamical perturbations.
According to \rfr{ierre}-\rfr{inorm} and \rfr{xyz}, the $R-T-N$ components of \rfr{sgorbia}
 are
 \begin{equation}
{\begin{array}{lll}
A_R^{(J_2)}& = &  -\rp{3 n^2  R_e^2 J_2}{8a(1-e^2)^4}\left(1+e\cos f\right)^4\left(1+3\cos 2\Psi +\right.\\ \\
&+& \left.6\sin^2\Psi \cos 2u\right), \\ \\
A_T^{(J_2)}& = & -\rp{3 n^2  R_e^2 J_2}{2a(1-e^2)^4}\left(1+e\cos f\right)^4\sin^2\Psi\sin 2u,\\ \\
A_N^{(J_2)}& = &   -\rp{3 n^2  R_e^2 J_2}{2a(1-e^2)^4}\left(1+e\cos f\right)^4\sin 2\Psi\sin u.
\end{array}}\lb{j2}
 \end{equation}
 Note that $[n^2 R^2_e a^{-1}]=$ L T$^{-2}$. For $\Psi=0$, i.e. for equatorial orbits, only the radial component is not zero. For polar orbits, i.e. for $\Psi=90$ deg, the normal component vanishes, contrary to the radial and transverse ones.
 \subsection{The effect of general relativity}
In its slow-motion and weak-field approximation, the Einstein's general theory of relativity predicts that a slowly rotating central body of mass $M$ and proper angular momentum $S$ induces two kinds of small perturbations on the orbital motion of a test particle. The largest one is dubbed gravitoelectric \citep{Mash}, and depends only on the mass $M$ of the body which acts as source of the gravitational field. It \textrm{is responsible of the} the well-known anomalous secular precession of the perihelion of Mercury of $43.98$ arcsec cty$^{-1}$ in the field of the Sun. There is also a smaller perturbation, known as gravitomagnetic \citep{Mash}, which depends on the angular momentum $S$ of the central body: it \textrm{causes} the Lense-Thirring \citep{LT} precessions of the node and pericenter of a test particle.
\subsubsection{The gravitoelectric, Schwarzschild-like perturbation}
By defining
\eqi \mathcal{R}_g\doteq \rp{G(M_{\star}+m_{\rm p})}{c^2}, \eqf where $c$ is the speed of light in vacuum, the $R-T-N$ components of the  general relativistic gravitoelectric perturbing acceleration are \citep{Soffel}
 \begin{equation}
{\begin{array}{lll}
A_R^{(\rm GE)}&= & \rp{n^2 \mathcal{R}_g}{(1-e^2)^3}(1+e\cos f)^2\left(3+2e\cos f -e^2 +\right.\\ \\
&+&\left. 4e^2\sin^2 f\right), \\ \\
A_T^{(\rm GE)}&= & \rp{n^2 \mathcal{R}_g}{(1-e^2)^3}(1+e\cos f)^2 4e\sin f(1+e\cos f),\\ \\
A_N^{(\rm GE)}&=& 0.
\end{array}}\lb{ge}
 \end{equation}
 Note that $[\mathcal{R}_g]=$ L, so that $[n^2 \mathcal{R}_g]=$ L T$^{-2}$.
\subsubsection{The gravitomagnetic, Lense-Thirring-like perturbation}
The $R-T-N$ components of the general relativistic gravitomagnetic perturbing acceleration induced by the rotation of the central body with proper angular momentum $S$ are \citep{Soffel}
 \begin{equation}
{\begin{array}{lll}
A_R^{(\rm GM)}&= & \eta_g\cos\Psi(1+e\cos f), \\ \\
A_T^{(\rm GM)}&= & -\eta_g e\cos\Psi\sin f,\\ \\
A_N^{(\rm GM)}&=& \eta_g\sin\Psi(1+e\cos f)\left[2\sin u + \right. \\ \\
&+& \left.e\left(\rp{\sin f\cos u}{1+e\cos f}\right)\right],
\end{array}}\lb{gm}
 \end{equation}
 with
 \eqi\eta_g\doteq\rp{\chi_g n}{a^2(1-e^2)^{7/2}}(1+e\cos f)^3,\eqf and \eqi \chi_g\doteq \rp{2 G S}{c^2}.\eqf
 Note that $[\chi_g]=$ L$^3$ T$^{-1}$, so that $[\eta_g]=$ L T$^{-2}$. For equatorial orbits $A_N^{\rm (GM)}=0$ and $A_R^{\rm (GM)}\neq 0, A_T^{\rm (GM)}\neq 0$. Instead, for polar orbits, i.e. for $\Psi=90$ deg, only the normal component does not vanish.
 \subsection{The tidal bulges}
 By neglecting the lag due to dissipation which is negligible for giant planets \citep{Murray}, the perturbing acceleration due to the tidal bulge raised on the planet by the host star is entirely radial, so that \citep{Cowling}
\begin{equation}
{\begin{array}{lll}
A_R^{(\rm tid\ p)}&= & -\left(\rp{M_{\star}}{m_{\rm p}}\right)\rp{ 3k_{2{\rm p}} r^5_{\rm p} G(M_{\star}+m_{\rm p})  }{r^7}, \\ \\
A_T^{(\rm tid\ p)}&= & 0,\\ \\
A_N^{(\rm tid\ p)}&=& 0.
\end{array}}\lb{tidacc}
 \end{equation}
 Here $k_{2{\rm p}}$ is the Love number of the planet\footnote{Here $k_{2{\rm p}}$ is twice the Love number customarily used in binary stars literature \citep{Cowling}.}. It  measures  how the mass redistribution induced  by the non-uniform stellar potential actually affects the external gravity field of the planet. Its values are in the range $k_{2{\rm p}}\approx 0.1-0.6$ \citep{Ragozzine}.

 A similar expression holds for the effect due to the tidal bulge raised by the planet on the star: the required substitutions are $M_{\star}\rightarrow m_{\rm p}, m_{\rm p}\rightarrow M_{\star}, r_{\rm p}\rightarrow R_{\star}, k_{2{\rm p}}\rightarrow k_{2\star}$.  Bodies with most of their mass near their cores, like main-sequence stars,  have very low $k_{2\star}$. It is so because the distorted outer envelope has little mass, thus affecting negligibly the outer gravitational field. Indeed, \citet{Claret} yields $k_{2\star}\approx 0.03$.
\subsection{A distant third body X}
 The perturbing potential induced by a distant, third body X is \citep{Ux}
\eqi U_{\rm X}=\rp{Gm_{\rm X}}{2r^3_{\rm X}}\left[r^2-3(\bds r\bds\cdot \bds{\hat{l}}_{\rm X})^2\right],\lb{UX}\eqf
where  $\bds{\hat{l}}_{\rm X}\doteq \bds{r}_{\rm X}/r_{\rm X}$ is a unit vector pointing towards X.
By denoting $l_x,l_y,l_z$ the direction cosines of $\bds{r}_X$, i.e. the components of  $\bds{\hat{l}}_{\rm X}$, we will express $U_{\rm X}$ as
\eqi U_{\rm X}=\rp{Gm_{\rm X}}{2r^3_{\rm X}}\left[r^2-3(xl_x+yl_y+zl_z)^2\right].\lb{UXnew}\eqf
After working out the cartesian components of the perturbing acceleration $\bds{A}_{\rm X}=-\bds{\nabla}U_{\rm X}$ from \rfr{UXnew}, the explicit expressions of the $R-T-N$ components can be worked out with the aid of \rfr{ierre}-\rfr{inorm} and \rfr{xyz}.
 They are
 \begin{equation}
{\begin{array}{lll}
A_R^{\rm (X)}&=& \rp{3a\left(1-e^2\right)\mathcal{K}_{\rm X}}{\left(1+e\cos f\right)}\left[\mathfrak{R}_s (\Psi_{\star},\Omega,\bds{\hat{l}}_{\rm X})\sin u +\right. \\ \\
 &+&\left.\mathfrak{R}_{c} (\Omega,\bds{\hat{l}}_{\rm X})\cos u\right]^2,\\ \\
A_T^{\rm (X)}&=& \rp{3a\left(1-e^2\right)\mathcal{K}_{\rm X}}{\left(1+e\cos f\right)}\left[\mathfrak{T}_s(\Psi_{\star},\Omega,\bds{\hat{l}}_{\rm X})\sin 2u +\right. \\ \\
&+&\left.\mathfrak{T}_c(\Psi_{\star},\Omega,\bds{\hat{l}}_{\rm X})\cos 2 u\right],\\ \\
A_N^{\rm (X)}&=& \rp{3a\left(1-e^2\right)\mathcal{K}_{\rm X}}{\left(1+e\cos f\right)}\left[\mathfrak{N}_s (\Psi_{\star},\Omega,\bds{\hat{l}}_{\rm X})\sin u +\right. \\ \\
&+&\left. \mathfrak{N}_c (\Psi_{\star},\Omega,\bds{\hat{l}}_{\rm X})\cos u\right],
\end{array}}\lb{accX}
 \end{equation}
 where
 \eqi \mathcal{K}_{\rm X}\doteq \rp{Gm_{\rm X}}{r^3_{\rm X}}\eqf is the so-called tidal parameter; $[\mathcal{K}_{\rm X}]={\rm T}^{-2}$. Note that both $\bds{\hat{l}}_{\rm X}$, which is assumed constant over one orbital revolution of the perturbed planet, and $\Omega$ here refer  to the stellar equatorial plane.
 %Actually, this matter depends on the context.
 %If the plane of the sky is chosen as reference $\{xy\}$ plane so that the $z$ axis is the line of sight, then $\bds{\hat{l}}_{\rm X}$ and\footnote{In this %case, the node is an angle in the plane of the sky from the axis pointing towards the North Celestial Pole to the intersection of the orbital plane to the %plane of the sky.} $\Omega$ are to be intended as referred to it,  and $\Psi\rightarrow i$.

The coefficients of \rfr{accX} are
%\eqi \mathfrak{R}_s\doteq 3\left(l_x\cos\Omega+l_y\sin\Omega\right)\left[l_z\sin\Psi +\cos\Psi\left(l_y\cos\Omega-l_x\sin\Omega\right)\right],\eqf
\begin{equation}
{
\begin{array}{lll}
\mathfrak{R}_{s}&\doteq & l_z\sin\Psi_{\star} +\cos\Psi_{\star}\left( l_y\cos\Omega-l_x\sin\Omega \right),\\ \\
\mathfrak{R}_{c} &\doteq & l_x\cos\Omega + l_y\sin\Omega,\\ \\
 \mathfrak{T}_{s}&\doteq & \rp{\left[l_z\sin\Psi_{\star}+\cos\Psi_{\star}\left(l_y\cos\Omega-l_x\sin\Omega\right)\right]^2-\left(l_x\cos\Omega+l_y\sin\Omega\right)^2}{2},\\ \\
\mathfrak{T}_{c}&\doteq & \left(l_x\cos\Omega+l_y\sin\Omega\right)\left[l_z\sin\Psi_{\star}+\cos\Psi_{\star}\left(l_y\cos\Omega-\right.\right.\\ \\
&-&\left.\left. l_x\sin\Omega\right)\right], \\ \\
 \mathfrak{N}_s &\doteq &\left[l_z\sin\Psi_{\star}+\cos\Psi_{\star}\left(l_y\cos\Omega-l_x\sin\Omega\right) \right]\times \\ \\
 &\times & \left[l_z\cos\Psi_{\star}+\sin\Psi_{\star}\left(l_x\sin\Omega- l_y\cos\Omega\right)\right], \\ \\
 \mathfrak{N}_c &\doteq & \left(l_x\cos\Omega+l_y\sin\Omega\right)\left[l_z\cos\Psi_{\star}+\sin\Psi_{\star}\left(l_x\sin\Omega-\right.\right. \\ \\
 &-&\left.\left. l_y\cos\Omega\right)\right].
 \end{array}
 }\lb{ruttu2_e}
 \end{equation}
 Note that, according to \rfr{ruttu2_e}, \rfr{accX} vanishes neither for $\Psi_{\star}=0$ nor for $l_z=0$. If both the planet and the perturber X are coplanar and lie in the equatorial plane of the star, then $A_N^{\rm (X)}=0$, while $A_R^{\rm (X)}\neq 0, A_T^{\rm (X)}\neq 0, A_R^{\rm (X)}\neq A_T^{\rm (X)}$, as intuitively expected.

 Actually, in view of their complexity, the previous formulas \textrm{do} not allow \textrm{one} to obtain manageable analytic expressions of general validity for the averaged time variations of the observable quantities we are interested in.
 Thus, in the following, we will release approximate expressions for such kind of perturbation.
 However, the exact effects of X can be numerically computed in specific arbitrary exoplanetary scenarios following the procedure discussed in Section \ref{intro}.
\section{Time variations of the transit duration in elliptic orbits}\lb{durazza}
For an unperturbed Keplerian elliptic orbit arbitrarily inclined to the line of sight, the transit duration $\Delta t_d$ can be written as \citep{Ting,Jordan}
\eqi \Delta t_d=\rp{2(R_{\star}+r_{\rm p})}{v}\cos\delta.\lb{trans}\eqf In it, $R_{\star}$ and $r_{\rm p}$ are the radii of the star and the planet, respectively. The Keplerian speed of the planet $v$ is \citep{Roy}
\eqi v=\rp{na}{\sqrt{1-e^2}}\sqrt{1+2e\cos f + e^2}.\lb{velka}\eqf  The latitude $\delta$ of the transit on the stellar disk is defined from
%\footnote{While \citet{Deeg}-see his Figure 2-neglects the planet's radius, \citet{Miralda} takes it into account. Note that \citet{Miralda} uses the angle %between the orbital plane and the line of sight, and denotes it as $\alpha$; in terms of our $i$, it is $\alpha=\pi/2+i$, so that $\sin\alpha\rightarrow \cos %i$. Moreover,  \citet{Miralda} adopts the letter $\gamma$ for the latitude of the transit $\delta$. Note that both \citet{Deeg} and \citet{Miralda} use $a$ %instead of $r$ in $\sin\delta$.} \citep{Ting,Jordan}
\eqi \sin\delta\doteq\rp{r\cos i}{R_{\star}+r_{\rm p}},\lb{trans2}\eqf in which
$r$ is given by \rfr{rkep}.
The parameter $i$ is the angle between the unit vector $\bds{\hat{L}}$ of the orbital angular momentum and the unit vector $\bds{\hat{\rho}}$ of the line of sight pointing towards the observer. From
the spherical law of cosines
%\footnote{See also http://mathworld.wolfram.com/SphericalTrigonometry.html on the WEB.}
\citep{Gel,Zwi}
\eqi\cos B=\sin C\sin A\cos b-\cos C\cos A\lb{trigga}\eqf  with the identifications\footnote{In such a way,  $\Omega$ results to be prograde with respect to the orbital motion, i.e. $\Omega$ follows it, coherently with the definition of the longitude of ascending node.} $A\rightarrow \Psi_{\star},B\rightarrow \pi-i,C\rightarrow I_{\star},b\rightarrow\pi-\Omega$,
 it turns out
 %\footnote{\citet{Miralda} denotes $\beta$ the angle between his mean plane-which in our case coincides with the star's equator-and the line of sight, so that %$\beta=\pi/2-I_{\star}$; with this change, eq. (9) and eq. (12) of \citet{Miralda} agrees with our \rfr{spherical} and \rfr{fina}. It must also be noted that %eq. (4) by \citet{Miralda} tells us that his $i_p$ and $i$ coincide with our $\Psi_{\star}$.}
\eqi\cos i=\sin\Psi_{\star}\sin I_{\star}\cos\Omega+\cos\Psi_{\star}\cos I_{\star},\lb{spherical}\eqf
where $\Psi_{\star}$ is the angle between $\bds{\hat{L}}$ and the unit vector $\bds{\hat{S}}_{\star}$ of the star's proper angular momentum, $I_{\star}$ is the angle between $\bds{\hat{S}}_{\star}$  and $\bds{\hat{\rho}}$, and $\Omega$ is the longitude of the ascending node defined from
\eqi\sin\Psi_{\star}\sin I_{\star}\cos\Omega
%=\bds{\hat{\tau}}\bds\cdot\bds{\hat{x}}
=(\bds{\hat{S}}_{\star} \bds\times \bds{\hat{L}})
\bds\cdot
(\bds{\hat{S}}_{\star} \bds\times \bds{\hat{\rho}}).\lb{megas}\eqf
Incidentally, it may be noted from \rfr{spherical} that for equatorial orbits, i.e. for $\Psi_{\star}=0$, we have $i=I_{\star}$.
Thus, $\Delta t_d$, for an elliptic orbit, is a function of\footnote{We neglect possible time changes of the star's spin axes.} $f,a,e$ through $v$ and $r$ in $\cos\delta$, and of $\Psi_{\star}$ and $\Omega$ through $i$ in $\cos\delta$. As a consequence, when the effect of a given dynamical perturbation on $\Delta t_d$ is considered, the variations of all such Keplerian orbital elements have to be fully taken into account.

In the following calculations of the time variation of $\Delta t_d$ due to various dynamical perturbations the expression \eqi \left\{
1-\left[\rp{a(1-e^2)\cos i}{(R_{\star}+r_{\rm p})(1+e\cos f)} \right]^2\right\}^{-1/2}\eqf appears: to make computation simpler, we will approximate it with
\eqi  \left\{1-\left[\rp{a(1-e^2)\cos i}{(R_{\star}+r_{\rm p})}\right]^2\right\}^{-1/2}.\lb{apra}\eqf
\subsection{The effect of the stellar oblateness}
\textrm{
An exact calculation, valid to all orders in $e$, cannot be performed because of the difficulty of the integrals involved in the evaluation of \rfr{ypa}.
Thus, in addition to \rfr{apra}, we have to approximate also \rfr{velka}  entering the integrands with negative powers. We will retain the first order terms in their expansions, which is admissible for $e<\sqrt{2}-1\approx 0.4$.}

\textrm{Such an approximate calculation} shows that the overall averaged time variation of $\Delta t_d$ induced by the stellar oblateness does not vanish for elliptic orbits. It turns out that the non-vanishing terms are those due to the perturbations in $a,e,\Psi_{\star},\Omega,\omega,\mathcal{\mathcal{M}}$. This is a clear example of how simply inserting the long-term precessions of the Keplerian orbital elements into instantaneous expressions would yield incorrect results; indeed, the long-term variations of $a,e,\Psi_{\star}$ caused by the oblateness of the primary are notoriously zero \citep{Roy}.
By defining
\eqi \sin\overline{\delta}\doteq \rp{a\cos i}{\left(R_{\star}+r_{\rm p}\right)},\eqf
we finally have
\eqi
\begin{array}{lll}
\left\langle\left.\dert{\Delta t_d}{t}\right|^{(J_2^{\star})}\right\rangle &=& -\left(\rp{R_{\star}}{a}\right)^2\rp{3 J_2^{\star}\sqrt{1-e^2}\sin\overline{\delta}  \sin 2\Psi_{\star}\sin I_{\star}\sin\Omega}{2\sqrt{1-(1-e^2)^2\sin^2\overline{\delta}}}+\\ \\
&+&\mathcal{O}(e^2).\lb{prizob}
\end{array}
\eqf
 In the limit $e\rightarrow 0$,
 \rfr{prizob} reduces to
  \eqi\left\langle\left.\dert{\Delta t_d}{t}\right|_0^{(J_2^{\star})}\right\rangle=\rp{2\tan\overline{\delta}\sin\Psi_{\star}\sin I_{\star}\sin\Omega}{n}\left\langle\left.\dert\Omega{t}\right|^{(J_2^{\star})}\right\rangle,\lb{prizok}\eqf
 obtained in \citet{Iorio010} for circular orbits.
 Indeed \citep{Roy},
 \eqi \left\langle\left.\dert\Omega{t}\right|_0^{(J_2^{\star})}\right\rangle=-\rp{3}{2}nJ^{\star}_2 \left(\rp{R_{\star}}{a}\right)^2\cos\Psi_{\star}\eqf
 for $e\rightarrow 0$. For another derivation of \rfr{prizok}, which can be applied to all perturbations when $e\rightarrow 0$, see Section \ref{icsdur}.
 Note that, for  exactly edge-on orbits,  both \rfr{prizob} and \rfr{prizok} vanish. The same occurs if the orbit is equatorial, irrespectively of its inclination  to the plane of the sky.
 \textrm{
 It is important to note that the long-term effects of \rfr{prizob}-\rfr{prizok} are not secular trends linearly changing in time because of the presence of $\Psi_{\star}$ and $\Omega$ as arguments of trigonometric functions. Indeed, such Keplerian orbital elements do, in general, slowly vary in time because of dynamical effects like a distant third body X, general relativity, and the oblateness itself.
 }

 Finally, let us mention that  the oblateness $j_2^{\rm p}$ of the planet \textrm{also} affects $\Delta t_d$ in the same way. The related formulas can be obtained from the previous one provided that the substitution $R_{\star}\rightarrow r_{\rm p}, J_2^{\star}\rightarrow j_2^{\rm p}$ is performed.
 %The angle $\psi_{\rm p}$ is the inclination of  the orbital plane to the equatorial plane of the planet.
\subsection{The effect of general relativity}
\subsubsection{The gravitoelectric, Schwarzschild-like perturbation}
 From \rfr{ge}  it turns out that the general relativistic gravitoelectric term can only affect $d\Delta t_d/dt$ through its perturbations on $a,e,\mathcal{M}$. The long-term variations for such Keplerian orbital elements are zero \citep{Soffel}, apart from the mean anomaly \citep{Iorio2}.

 Instead,  after an approximate  calculation valid for $e<\sqrt{2}-1\approx 0.4$ analogous to the previous one, it turns out that
 \eqi \left\langle\left.\dert{\Delta t_d}{t}\right|^{(\rm GE)}\right\rangle=0.\eqf

 The same holds also for circular orbits, as can be immediately seen by inspecting the general method of Section \ref{icsdur} which involves the long-term variations of $\Psi_{\star}$ and $\Omega$.
 %This is an exact result since no approximations in $e$ have been used.
\subsubsection{The gravitomagnetic, Lense-Thirring-like perturbation}
According to an approximate calculation in $e$, valid for $e<\sqrt{2}-1\approx 0.4$, analogous to the previous ones,  \textrm{the} averaged time variation of $\Delta t_d$ induced by the general relativistic gravitomagnetic effect is, in general, non-zero for elliptic orbits. In particular, the non-vanishing terms are those due to the perturbations in $\Psi_{\star}$ and $\Omega$. This is another  example of how the mere insertion of the long-term precessions of the Keplerian orbital elements into instantaneous expressions would yield incorrect results; indeed, the Lense-Thirring long-term precession of $\Psi_{\star}$ vanishes \citep{LT,Soffel}.

We have
\eqi
\begin{array}{lll}
\left\langle\left.\dert{\Delta t_d}{t}\right|^{\rm (GM)}\right\rangle & = & \left(\rp{GS_{\star}}{c^2 a^3 n}\right)\rp{\sin\overline{\delta}\sin\Psi_{\star}}{2 \sqrt{1-(1-e^2)^2\sin^2\overline{\delta}}}\times \\ \\
&\times &\left[\mathfrak{L}(e,\Omega,\omega,I_{\star},\Psi_{\star})+\mathfrak{M}(e,\Omega,\omega,I_{\star})\right],\lb{letazz}
\end{array}
\eqf
with
\begin{equation}
{\begin{array}{lll}
\mathfrak{L}&\doteq & e^2\left(1-e^4\right)\left(\cos I_{\star}\sin\Psi_{\star}-\right. \\ \\
&-&\left.\cos\Psi_{\star}\sin I_{\star}\cos\Omega\right)\sin 2\omega, \\ \\
\mathfrak{M}&\doteq & \left(1-e^2\right)\left\{8\left[1-e^2\left(1-\rp{e^2}{2}\right)\right]-\right. \\ \\
&-& \left.e^2\left(1+e^2\right)\cos 2\omega\right\}\sin I_{\star}\sin\Omega;
\end{array}}\lb{letizz}
 \end{equation}
 the term containing $\mathfrak{L}$ comes from the perturbation in $\Psi_{\star}$, while the term with $\mathfrak{M}$ is due to $d\Omega/dt$.

 For $e\rightarrow 0$, \rfr{letazz}-\rfr{letizz} yield \citep{Iorio010}
 \eqi
 \begin{array}{lll}
 \left\langle\left.\dert{\Delta t_d}{t}\right|_0^{\rm (GM)}\right\rangle & = & \rp{2 \tan\overline{\delta}\sin \Psi_{\star}\sin I_{\star}\sin\Omega}{n}\left(\rp{2 GS_{\star} }{c^2 a^3}\right)=\\ \\
 &=&\rp{2\tan\overline{\delta}\sin \Psi_{\star}\sin I_{\star}\sin\Omega}{n}\left\langle\left.\dert{\Omega}{t}\right|^{\rm (GM)}\right\rangle.\lb{letezz}
 \end{array}
 \eqf

 \textrm{
 According to \rfr{letazz}-\rfr{letezz}, also the long-term gravitomagnetic time variation of $\Delta t_d$ is a harmonic one.
 }
 For  exactly edge-on orbits,  both \rfr{letazz} and \rfr{letezz} vanish. The same occurs if the orbit is equatorial, \textrm{independent} of its orientation with respect to the line of sight.
\subsection{The tidal bulges}
It \textrm{is found} that the long-term variations of the duration transit caused by the tidal bulges mutually raised by the planet and the star on each other vanish.
\subsection{A distant third body X}\lb{icsdur}
It turns out that a distant body X induces a  long-term \textrm{harmonic} variation of the duration transit which does not vanish in the limit $e\rightarrow 0$.
In this case, we have
\eqi \left\langle\left.\dert{\Delta t_d}{t}\right|_0^{\rm (X)}\right\rangle=\left(\rp{\mathcal{K}_{\rm X}}{n^2}\right)3\tan\overline{\delta}\mathfrak{X}(\bds{\hat{l}}_{\rm X},\Psi_{\star},I_{\star},\Omega),\lb{xrate}\eqf
with
\begin{equation}
{\begin{array}{lll}
\mathfrak{X} &\doteq &\left[l_z \cos\Psi_{\star} +
    \sin\Psi_{\star} \left(-l_y \cos\Omega + l_x \sin\Omega\right)\right]\times\\ \\
    &\times &\left\{
    -l_x\cos\Psi_{\star}\sin I_{\star}+\sin\Psi_{\star}
    \left[
    l_z\sin I_{\star}\sin\Omega+\right.\right.\\ \\
    &+&\left.\left.\cos I_{\star}\left(l_x\cos\Omega+l_y\sin\Omega\right)
    \right]
    \right\}.
%l_x \sin I_{\star} \left(l_y \cos\Omega- l_x \sin\Omega\right)+l_z\left[l_z\sin I_{\star}\sin\Omega+\cos %I_{\star}\left(l_x\cos\Omega+l_y\sin\Omega\right)\right].
\end{array}}\lb{strazio}
 \end{equation}
 Note that if both the perturber body X and the perturbed planet lie in the equatorial plane of the host star, i.e. if $\Psi_{\star}=0$ and $l_z=0$, then
 \eqi\left\langle\left.\dert{\Delta t_d}{t}\right|_0^{\rm (X)}\right\rangle=0,\eqf
 independently of the inclination of the orbital plane to the plane of the sky. Instead, if $\Psi_{\star}=0$, but X is not coplanar with the perturbed planet, the induced long-term variation of the duration transit does not vanish.  The duration transit is not affected by X if the orbit of the perturbed planet is exactly edge-on.

 It is interesting to note that \rfr{xrate}, with \rfr{strazio}, can be also obtained in another, simpler way; the following approach is valid for all the dynamical perturbations previously examined. Indeed, for circular orbits the expression for $\Delta t_d$ of \rfr{trans} gets simplified in such a way that
 \eqi \left.\dert{\Delta t_d}{t}\right|_0=\rp{2\tan\overline{\delta}\sin i}{n}\left(\dert{i}{t}\right)+\derp{\Delta t_d}{a}\left(\dert{a}{t}\right).\lb{ehu}\eqf
 From \rfr{spherical} it turns out
 \eqi
 \begin{array}{lll}
 \sin i \left(\dert{i}{t}\right) & = & \left(\sin\Psi_{\star}\cos I_{\star}-\cos\Psi_{\star}\sin I_{\star}\cos\Omega\right)\dert{\Psi_{\star}}{t}+\\ \\
 &+& \sin\Psi_{\star}\sin I_{\star}\sin\Omega \left(\dert{\Omega}{t}\right).\lb{eho}
 \end{array}
 \eqf Thus, only for circular orbits it is admissible to straightforwardly use the long-term variations of $a,\Psi_{\star}$ and $\Omega$, as shown by \rfr{ehu}-\rfr{eho}.
 In the case of the perturbation due to a distant body X, it turns out from \rfr{Gauss}
 \begin{equation}
{\begin{array}{lll}
\left\langle\left.\dert{a}{t}\right|^{\rm (X)}\right\rangle &=& 0,\\ \\
\left\langle\left.\dert{\Psi_{\star}}{t}\right|^{\rm (X)}\right\rangle &=& \rp{3 \mathcal{K}_{\rm X}(1-e^2)^3}{2 n}\left(l_x\cos\Omega+l_y\sin\Omega\right)\times \\ \\
&\times &\left[l_z\cos\Psi_{\star}+\sin\Psi_{\star}\left(l_x\sin\Omega-l_y\cos\Omega\right)\right],\\ \\
\left\langle\left.\dert{\Omega}{t}\right|^{\rm (X)}\right\rangle &=& \rp{3 \mathcal{K}_{\rm X}\csc\Psi_{\star}(1-e^2)^3}{2 n}\left[l_z\sin\Psi_{\star}+\cos\Psi_{\star}\times \right.\\ \\
&\times & \left.\left(l_y\cos\Omega-l_x\sin\Omega\right)\right]\left[l_z\cos\Psi_{\star}+\right.\\ \\
&+&\left.\sin\Psi_{\star}\left(l_x\sin\Omega-l_y\cos\Omega\right)\right],
\end{array}}\lb{strazio2}
 \end{equation}
 contrary to the stellar oblateness and the general relativistic gravitomagnetism for which the long-term precessions of $\Psi_{\star}$ vanish.
 It turns out that \rfr{strazio2}, \textrm{substituted} in \rfr{ehu}-\rfr{eho}, yields the same result of \rfr{xrate}-\rfr{strazio}.
\subsection{Numerical evaluations}
A feature common to all the variations of $\Delta t_d$ previously examined is that they all vanish for exactly edge-on orbits. Moreover, independently of the inclination of the orbital plane to the line of sight, they vanish also for equatorial orbits, apart from the case of X unless it is coplanar with the perturbed planet moving in the star's equatorial plane.

Thus, we assume in the following $i=87$ deg and $\Psi_{\star}=15$ deg.
With such assumptions we have
\begin{equation}
{\begin{array}{lll}
\left|\left\langle\left.\dert{\Delta t_d}{t}\right|^{(j_2^{\rm p})}\right\rangle\right| &\lesssim & 1\times 10^{-8},\\ \\
\left|\left\langle\left.\dert{\Delta t_d}{t}\right|^{(J_2^{\star})}\right\rangle\right| &\lesssim & 9\times 10^{-10},\\ \\
\left|\left\langle\left.\dert{\Delta t_d}{t}\right|^{\rm (GM)}\right\rangle\right| &\lesssim & 1\times 10^{-12}, \\ \\
\left|\left\langle\left.\dert{\Delta t_d}{t}\right|^{\rm (tid\ \star)}\right\rangle\right| & = & 0\\ \\
\left|\left\langle\left.\dert{\Delta t_d}{t}\right|^{\rm(tid\ p)}\right\rangle\right| & = & 0,\\ \\
\left|\left\langle\left.\dert{\Delta t_d}{t}\right|^{\rm(GE)}\right\rangle\right| & = & 0.\\ \\
\end{array}}\lb{numeracciacci}
 \end{equation}
By assuming $e=0$, a third body X with the mass of Jupiter and located at $r_{\rm X}=0.1-1$ au yields
\eqi  \left|\left\langle\left.\dert{\Delta t_d}{t}\right|^{\rm(Jup)}\right\rangle\right| \lesssim 8\times 10^{-5}-8\times 10^{-8},\eqf
according to \ref{xrate}.
The effect of an Earth-like perturber  $r_{\rm X}=0.1-1$ au is, instead,
 \eqi  \left|\left\langle\left.\dert{\Delta t_d}{t}\right|^{\rm(Ear)}\right\rangle\right| \lesssim 2\times 10^{-7}-2\times 10^{-10}.\eqf

%It may be interesting to note that such figures remain substantially the same also for $\Psi_{\star}=45$ deg.
The time variations of the duration transit induced by the  planetary and stellar oblateness for closer orbits, i.e. for $a=0.015$ au, would be of the order of $7\times 10^{-7}\ (j_2^{\rm p})$ and $2\times 10^{-9}\ (J_2^{\star})$, respectively.
\textrm{In Figure \ref{figurona1} we plot the upper bounds in $\left\langle\dot \Delta t_{d}\right\rangle$ due to all the dynamical effects considered, apart from the third body X, as a function of $a$ for different values of the eccentricity. We let $a$ vary within the range of the so-far discovered transiting exoplanets\footnote{See on the WEB http://www.exoplanet.eu.}, with a maximum value of $e$ yielding  periastron distances not smaller than the star's radius. We adopted nearly edge-on ($i=87$ deg) and almost equatorial ($\Psi_{\star}=15$ deg) orbital configurations. The physical parameters of the star and the planet have been retrieved from Table \ref{parametri}.
\begin{figure*}
\centering
\begin{tabular}{ccc}
\epsfig{file=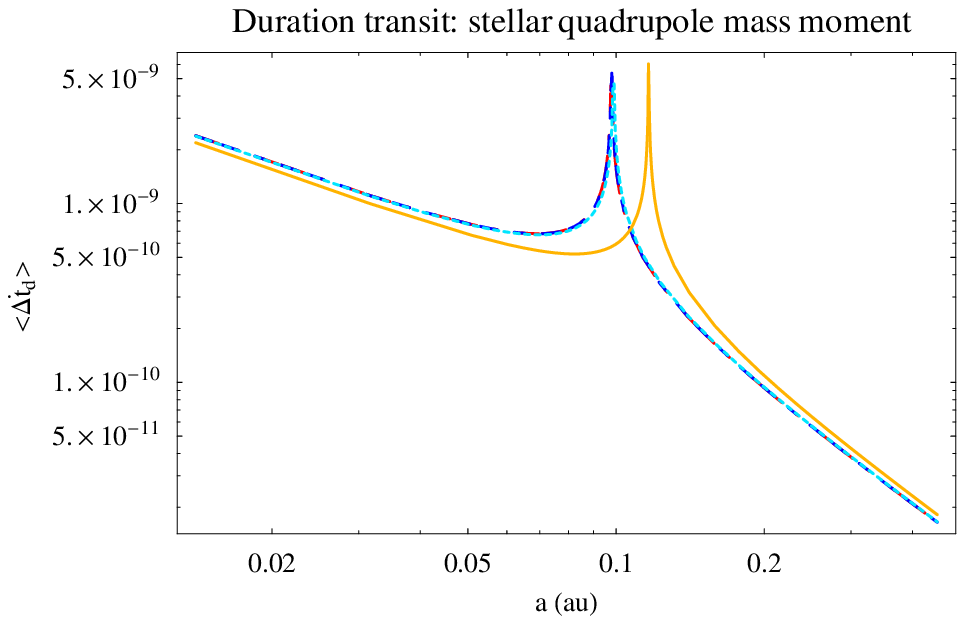,width=0.31\linewidth,clip=} &
\epsfig{file=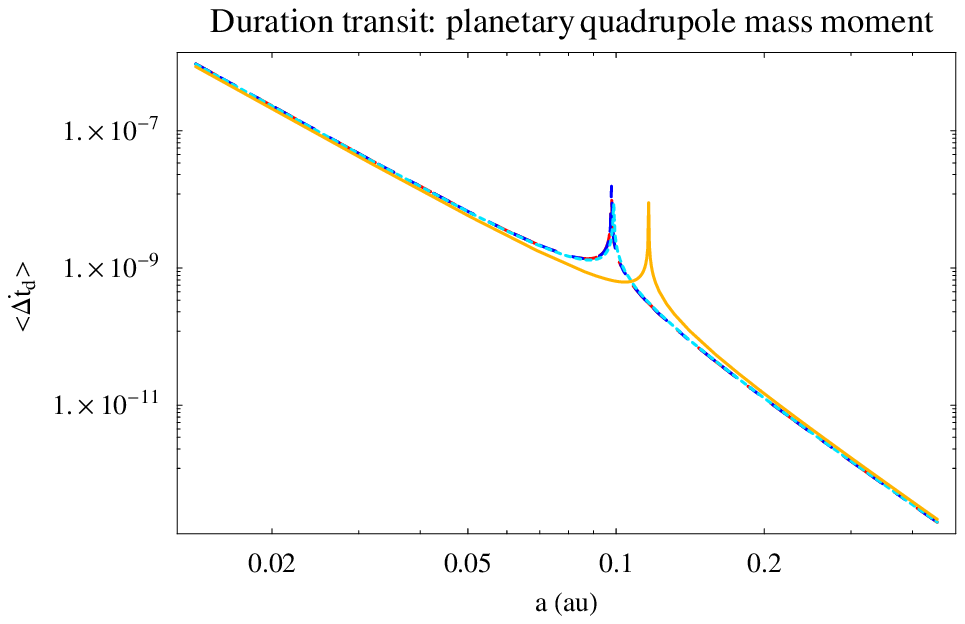,width=0.31\linewidth,clip=} &
\epsfig{file=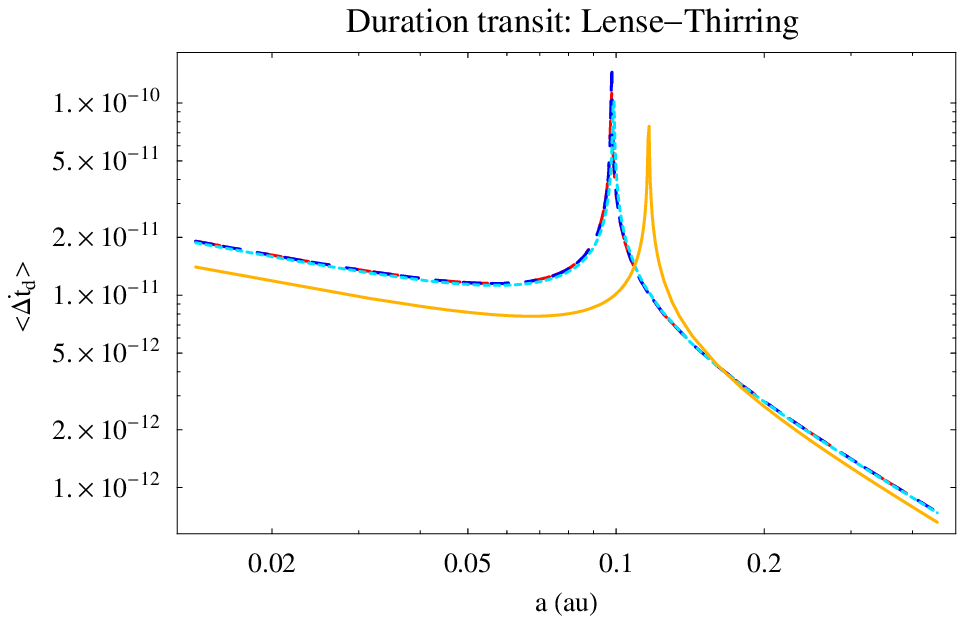,width=0.31\linewidth,clip=}
\end{tabular}
\caption{Maximum values of the long-term time variations $\left\langle\dot \Delta t_{d}\right\rangle$ as a function of $a$ ($0.014\ {\rm au}\leq a\leq 0.449\ {\rm au}$) for different values of the eccentricity: $e=0.005$ (red dash-dotted line), $e=0.03$ (blue dashed line), $e=0.1$ (light blue dotted line), $e=0.4$ (yellow continuous line). For the stellar and planetary physical parameters we  used the standard values of Table \ref{parametri}. We adopted $i=87$ deg, $\Psi_{\star}=15$ deg, while we fixed the periastron at $\omega=90$ deg; we also considered $\sin I_{\star}\sin\Omega\approx 1$. }\lb{figurona1}
\end{figure*}
It is interesting to look also at almost polar orbital configurations ($\Psi_{\star}\approx 90$ deg). This is done in Figure \ref{figurona1bis} obtained for the same values of Figure \ref{figurona1}, apart from the inclination of the orbital plane to the stellar equator set now to $\Psi_{\star}=89$ deg.
\begin{figure*}
\centering
\begin{tabular}{ccc}
\epsfig{file=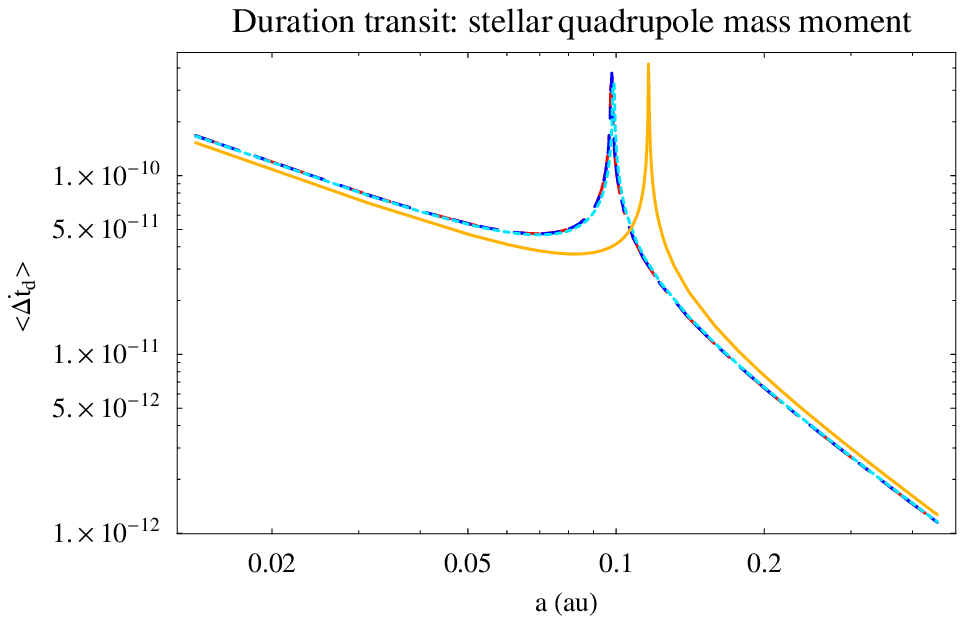,width=0.31\linewidth,clip=} &
\epsfig{file=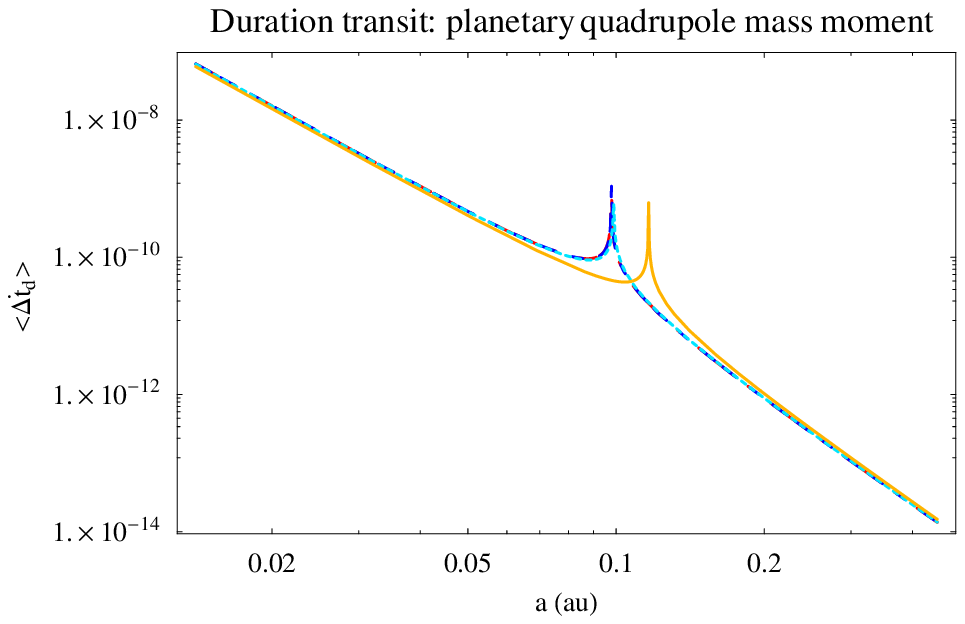,width=0.31\linewidth,clip=} &
\epsfig{file=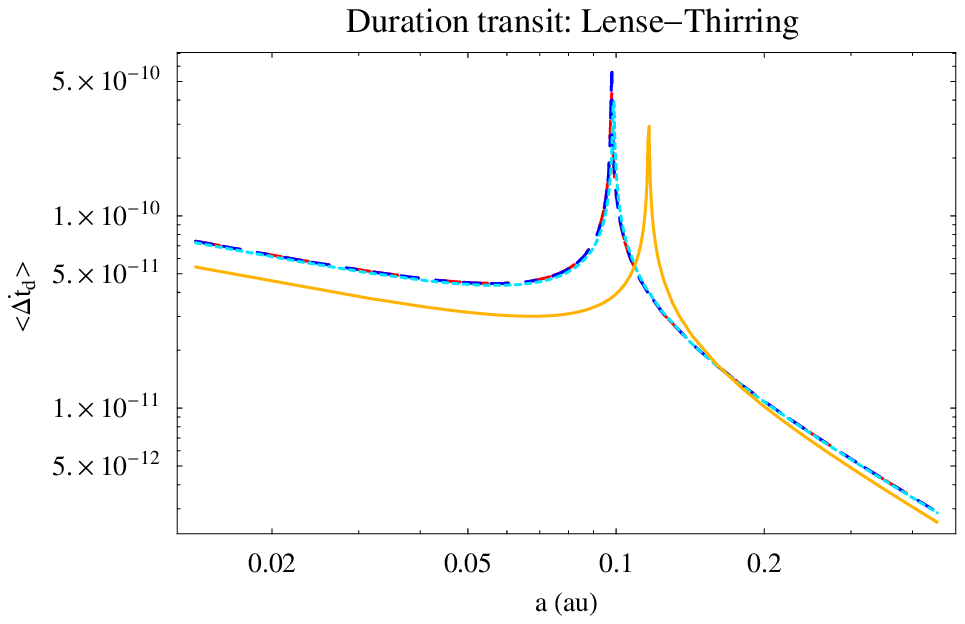,width=0.31\linewidth,clip=}
\end{tabular}
\caption{Maximum values of the long-term time variations $\left\langle\dot \Delta t_{d}\right\rangle$ as a function of $a$ ($0.014\ {\rm au}\leq a\leq 0.449\ {\rm au}$) for different values of the eccentricity: $e=0.005$ (red dash-dotted line), $e=0.03$ (blue dashed line), $e=0.1$ (light blue dotted line), $e=0.4$ (yellow continuous line). For the stellar and planetary physical parameters we  used the standard values of Table \ref{parametri}. We adopted $i=87$ deg, $\Psi_{\star}=89$ deg, while we fixed the periastron at $\omega=90$ deg; we also considered $\sin I_{\star}\sin\Omega\approx 1$. }\lb{figurona1bis}
\end{figure*}
}
%NearlyPolar orbits, i.e. $\Psi_{\star}=90$ deg,
 \subsection{The observability of the time variation of the transit duration}
 The accuracy obtainable in measuring the transit duration is of the order of $1.5-5$ s; see the accurate discussion in  \citet{Ford} and \citet{Jordan}. Thus, over an observational time span of $\tau=10$ yr it should be possible to achieve an accuracy in measuring the time variation of the transit duration of the order of
 \eqi\sigma_{\dot\Delta t_d}\approx 5\times 10^{-9}-1.6\times 10^{-8}.\eqf
 A similar conclusion is reached by \citet{Miralda}.
 Thus, it would be very difficult to measure the effects previously examined with the time variations of the duration transit in typical exoplanetary scenarios, with the possible exception of a third body X.
\section{The radial velocity}\lb{velazza}
The basic observable in spectroscopic studies of exoplanets, transiting or not,  is the radial velocity $V_{\rho}$.
Its expression for elliptic orbits, up to the velocity of the system's center of mass $V_0$, is \citep{Ency}
%can be obtained by using the\footnote{Recall that the reference $\{xy\}$ plane is the plane of the sky, so that the line of sight is the $z$ axis} $z$ %components of \rfr{ierre} and \rfr{itrav}, replacing $\Psi_{\star}$ with $i$ in them, and recalling that, for a Keplerian orbit,
%\eqi \bds v=v_R\bds{\hat{R}}+v_T\bds{\hat{T}}=\rp{na}{\sqrt{1-e^2}}\left[e\sin f\bds{\hat{R}}+(1+e\cos f)\bds{\hat{T}}\right].\eqf The result is\footnote{Also %the radial component of the velocity of the center of mass of the system, usually dubbed $V_0$, is added to \rfr{radvel}.}
\eqi V_{\rho}=K\left[e\cos\omega+\cos(f+\omega)\right],\lb{radvel}\eqf
where $2K$ is the total observed range of radial velocity
defined by
\eqi K\doteq\rp{na\sin i}{\sqrt{1-e^2}}.\lb{kvel}\eqf

Perturbing dynamical effects affect the radial velocity as well by inducing, in principle, a non-vanishing  net radial acceleration over one orbital period.
It can straightforwardly be worked out from \rfr{ypa} with $Y\rightarrow V_{\rho}$ by noting that, in this case,  the perturbations of all the six Keplerian orbital elements are involved.
\subsection{The effect of the stellar oblateness}
The stellar oblateness causes a long-term \textrm{harmonic} variation of the radial velocity only if the orbit is elliptic. Indeed, it turns out
\eqi
\begin{array}{lll}
\left\langle\dot V_{\rho}^{ (J_2^{\star})}\right\rangle & = & -\left(\rp{n^2 R^2_{\star}}{a}\right)\rp{3e  J_2^{\star} }{32 \left(1-e^2\right)^{7/2} \sin i}\times \\ \\
&\times & \left[\mathcal{J}_c(e,i,I_{\star},\Omega,\Psi_{\star})\cos\omega+\right. \\ \\
& + &\left.\mathcal{J}_s(e,i,I_{\star},\omega,\Omega,\Psi_{\star})\sin\omega\right],\lb{velj2}
\end{array}
\eqf
with
\begin{equation}
{\begin{array}{lll}
\mathcal{J}_c &\doteq & 10(1-e^2)\cos i \sin I_{\star}\sin 2 \Psi_{\star}\sin\Omega,\\ \\
\mathcal{J}_s &\doteq & 2\left(1-e^2\right)\cos i\sin 2\Psi_{\star}\left(\cos I_{\star}\sin\Psi_{\star}-
\right.\\ \\
&-&\left.\sin I_{\star}\cos\Psi_{\star}\cos\Omega\right)+ \sin^2 i \left[7+47\cos 2\Psi+\right. \\ \\
&+&\left.\sin^2\Psi_{\star}\cos 2\omega-\rp{e^2}{16} \left(259+429\cos 2\Psi_{\star}-
\right.\right.\\ \\
&-&\left.\left. 44\sin^2\Psi_{\star}\cos 2\omega\right)\right].
\end{array}}\lb{coeffvelj2}
 \end{equation}
 It is an exact result in $e$, and vanishes in the limit $e\rightarrow 0$. Note that, according to \rfr{coeffvelj2}, \rfr{velj2} vanishes neither for equatorial orbits nor for edge-on configurations.
\subsection{The effect of general relativity}
Also general relativity affects the radial velocity of non-circular orbits. Indeed, an exact calculation in $e$ yields \textrm{the following long-term harmonic signatures.}
\eqi\left\langle\dot V_{\rho}^{\rm (GE)}\right\rangle = \left(n^2 {\mathcal{R}_g}\right)\rp{ 15 e(1+e^2)\sin i \sin\omega}{8\left(1-e^2\right)^{5/2}},\lb{velge}\eqf
and
\eqi
\begin{array}{lll}
\left\langle\dot V_{\rho}^{\rm (GM)}\right\rangle &=& \left(\rp{nGS_{\star}}{c^2 a^2}\right)\rp{e}{4 (1-e^2)^{2}}\left[\mathcal{V}_c(i,I_{\star},\Omega,\Psi_{\star})\cos\omega+\right. \\ \\
&+&\left.\mathcal{V}_s(i,I_{\star},\Omega,\Psi_{\star})\sin\omega\right],\lb{velgm}
\end{array}
\eqf
with
\begin{equation}
{\begin{array}{lll}
\mathcal{V}_c &\doteq & 11\cot i \sin I_{\star}\sin\Psi_{\star}\sin\Omega,\\ \\
\mathcal{V}_s &\doteq &\rp{\csc i}{4}\left\{\cos\Omega\sin 2I_{\star}\left(\sin\Psi_{\star} -\sin 3\Psi_{\star}\right)
-\right.\\ \\
&-& \left. \sin^2\Psi_{\star}\cos\Psi_{\star}\left[\cos 2I_{\star} \left(3 + \cos 2\Omega\right) +
        2\sin^2\Omega\right]+ \right.\\ \\
        &+&\left. 104\sin^2 i \cos\Psi_{\star}
\right\}.
\end{array}}\lb{coeffvelgm}
 \end{equation}
For $e\rightarrow 0$ both \rfr{velge} and \rfr{velgm} vanish.
Also the general relativistic effects are non-vanishing either for edge-on or equatorial orbits, with \rfr{velge} which is independent of $\Psi_{\star}$, contrary to \rfr{velgm}.
\subsection{The tidal bulges}\lb{sgulla}
The tidal bulge of the planet induces a non-zero long-term \textrm{harmonic} variation of the radial velocity. Its exact expression is
\eqi
\begin{array}{lll}
&\left\langle \dot V_{\rho}^{\rm (tid\ p)}\right\rangle & =\\ \\
 & =&\left(\rp{n^2 r_{\rm p}^5}{a^4}\right)\left(\rp{M_{\star}}{m_{\rm p}}\right)\times \\ \\
 &\times &\rp{3e  \left[-496+5e^2\left(-80+81 e^2+8e^4\right)\right]k_{2{\rm p}} \sin i \sin\omega }{128  (1-e^2)^{13/2}}.
\lb{trimma}
\end{array}
\eqf
The long-term variation of the radial velocity due to the tidal bulge raised by the planet on the star is
\eqi
\begin{array}{lll}
&\left\langle \dot V_{\rho}^{\rm (tid\ \star)}\right\rangle & = \\ \\
 & =&\left(\rp{n^2 R_{\star}^5}{a^4}\right)\left(\rp{m_{\rm p}}{M_{\star}}\right)\times \\ \\
 &\times &\rp{3e   \left[-496+5e^2\left(-80+81 e^2+8e^4\right)\right]k_{2{\star}} \sin i \sin\omega }{128  (1-e^2)^{13/2}}.
\lb{tramma}\end{array}
\eqf
They vanish for $e\rightarrow 0$ and are independent of $\Psi_{\star}$.

\textrm{It is easy to understand that, concerning the tidal distortions, those experienced by the planet are more effective than those suffered by its hosting star. Indeed, \rfr{trimma}-\rfr{tramma} tell us that their ratio goes as
\eqi \left(\rp{M_{\star}}{m_{\rm p}}\right)^2\left(\rp{r_{\rm p}}{R_{\star}}\right)^5\rp{k_{\rm 2p}}{k_{2\star}}.\lb{rakka}\eqf
According to Table \ref{parametri}, the ratio of the star-planet masses is typically of the order of $10^3$, while the planet-to-star ratios for the radii and the Love numbers are about $0.1$ and $20$, respectively.
}
\subsection{A distant third body X}
 A net non-zero \textrm{long-term harmonic} effect on $V_{\rho}$ is induced by a distant third body X. Its general expression for arbitrary values of the eccentricity and of the position of X is too cumbersome to be explicitly displayed here; it turns out
 \eqi \left\langle \dot V_{\rho}^{\rm (X)}\right\rangle\propto -\left(a\mathcal{K}_{\rm X}\right) \rp{3e(1-e^2)^{3/2}}{8\sin i}\mathfrak{W}(\Omega,\omega,\Psi_{\star},\bds{\hat{l}}_{\rm X}).\lb{zomba}\eqf
 The quite complicated function $\mathfrak{W}$ does not vanish for $l_z=0,\Psi_{\star}=0$, i.e. if both the perturbed and the perturbing planets lie in the star's equatorial plane.
 As for the stellar oblateness and general relativity, $\left\langle\dot V_{\rho}^{\rm (X)}\right\rangle$  vanishes for $e\rightarrow 0$. It should not be considered as contradictory because it is the otherwise constant velocity $V_0$ of the system's center of mass that is altered by the perturber X, independently of the relative  motion of the perturbed planet and the star.

\subsection{Numerical evaluations}
The \virg{standard} orbital scenario of Table \ref{parametri} yields
\begin{equation}
{\begin{array}{lll}
\left|\left\langle\dot V_{\rho}^{\rm(tid\ p)}\right\rangle\right| &\leq & 4\times 10^{-7} \ {\rm m\ s}^{-2},\\ \\
\left|\left\langle\dot V_{\rho}^{\rm(GE)}\right\rangle\right| &\leq & 1\times 10^{-7}\ {\rm m\ s}^{-2},\\ \\
\left|\left\langle\dot V_{\rho}^{(j_2^{\rm p})}\right\rangle\right| &\leq & 6\times 10^{-8}\ {\rm m\ s}^{-2},\\ \\
\left|\left\langle\dot V_{\rho}^{(J_2^{\star})}\right\rangle\right| &\leq & 3\times 10^{-9}\ {\rm m\ s}^{-2},\\ \\
\left|\left\langle\dot V_{\rho}^{\rm (tid\ \star)}\right\rangle\right| &\leq & 2\times 10^{-9}\ {\rm m\ s}^{-2},\\ \\
\left|\left\langle\dot V_{\rho}^{\rm (GM)}\right\rangle\right| &\leq & 9\times 10^{-11}\ {\rm m\ s}^{-2}. \\ \\
\end{array}}\lb{numerelli}
 \end{equation}
Concerning a distant third body X, for $m_{\rm X}=m_{\rm Jup}, r_{\rm X}=0.1-1$ au and $l_z=0$, i.e by assuming coplanarity with the perturbed planet, we have
\eqi \left|\left\langle\dot V_{\rho}^{\rm (Jup)}\right\rangle\right|\lesssim 1\times 10^{-5}-1\times 10^{-8}\ {\rm m\ s}^{-2}. \eqf
For an Earth-sized perturber at $0.1-1$ au we have
\eqi
\left|\left\langle\dot V_{\rho}^{\rm (Ear)}\right\rangle\right|\lesssim 5\times 10^{-8}-5\times 10^{-11}. \eqf

\textrm{
In Figure \ref{figurona2} we plot the magnitude of the upper bound of $\left\langle\dot V_{\rho}\right\rangle$ for all the dynamical effects considered, apart from a third body X, as a function of the semi-major axis for different values of the eccentricity. Also in this case, we let $a$ vary within the range of the so-far discovered transiting exoplanets\footnote{See on the WEB http://www.exoplanet.eu.}, with a maximum value of $e$ yielding  periastron distances not smaller than the star's radius. Table \ref{parametri} has been used for the values of the physical parameters of the star and the planet, and for the various orbital inclinations $i,\Psi_{\star}, \psi_{\rm p}$; the periastron has been set equal to $\omega=90$ deg.
\begin{figure*}
\centering
\begin{tabular}{cc}
\epsfig{file=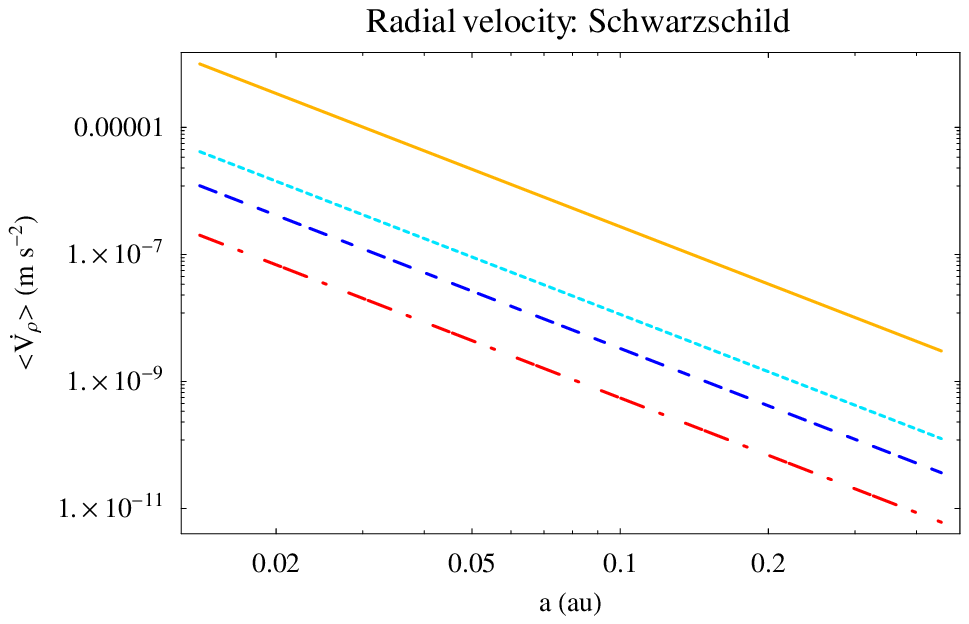,width=0.45\linewidth,clip=} &
\epsfig{file=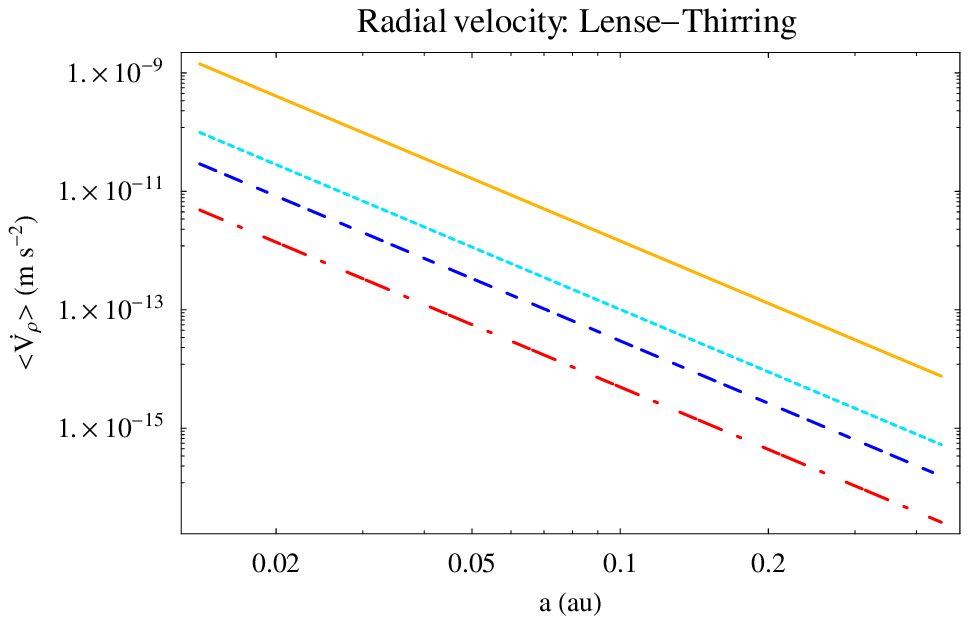,width=0.45\linewidth,clip=} \\
\epsfig{file=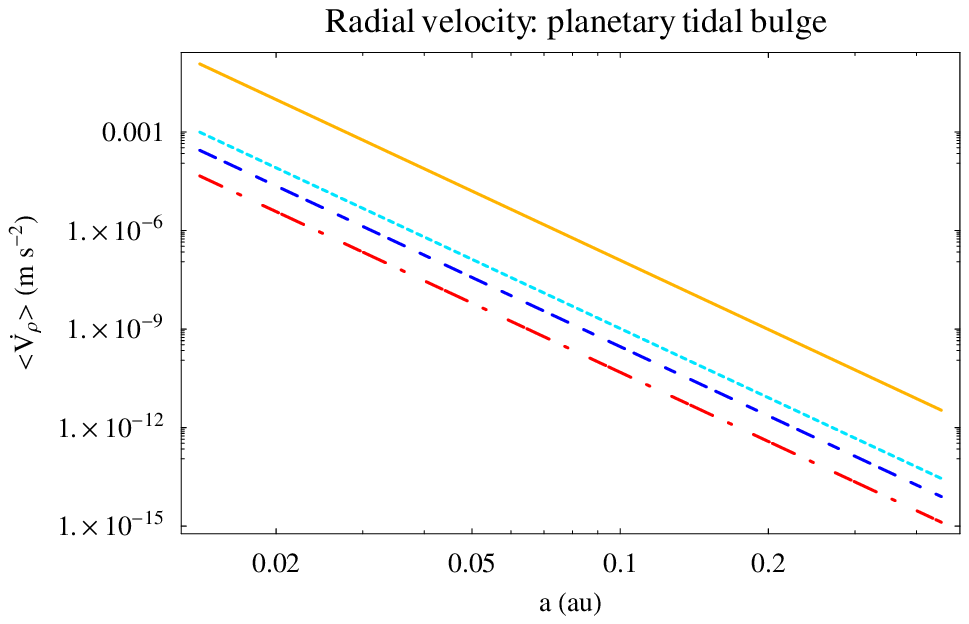,width=0.45\linewidth,clip=} &
\epsfig{file=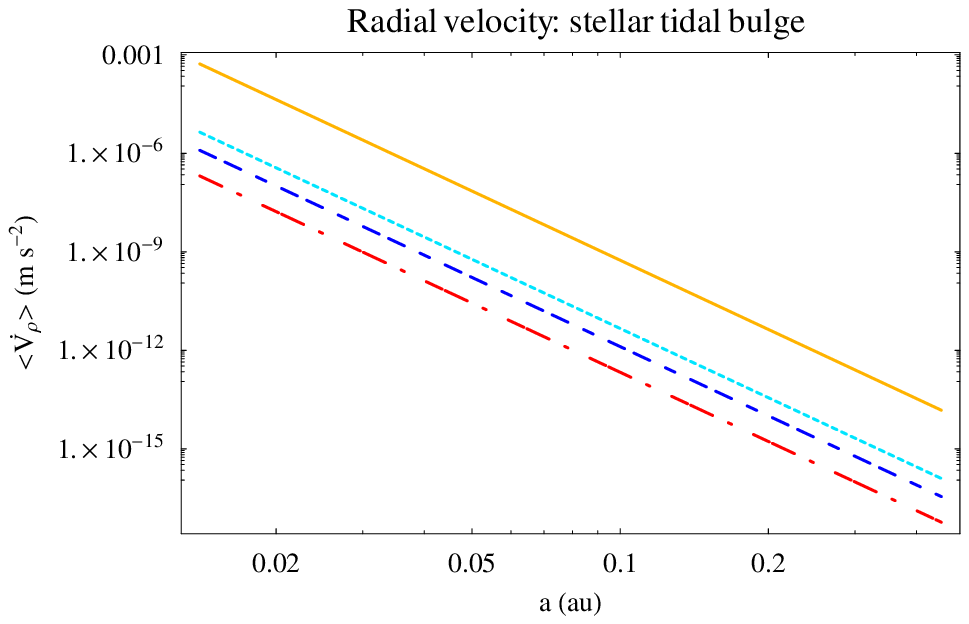,width=0.45\linewidth,clip=}\\
\epsfig{file=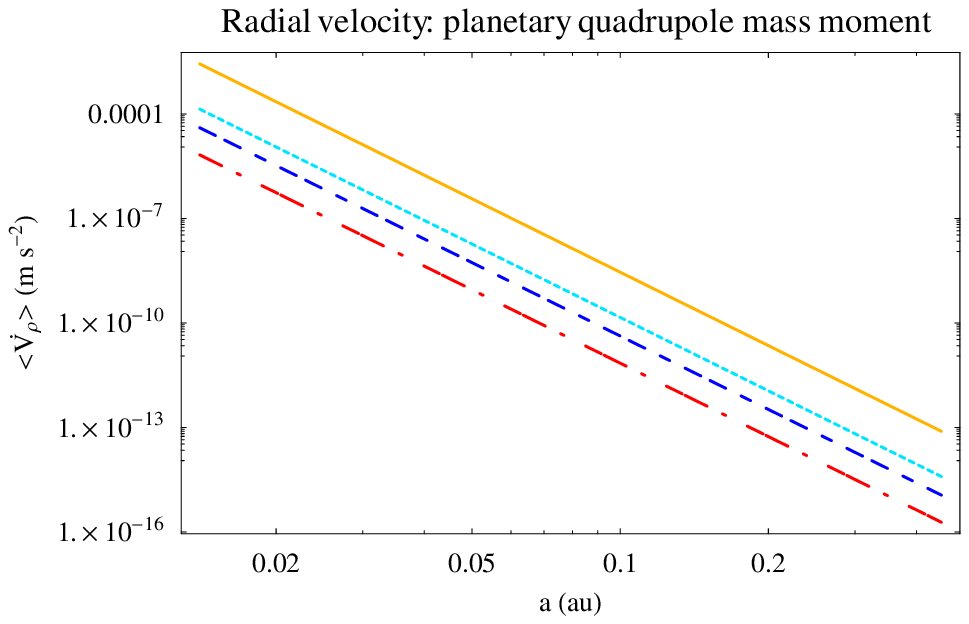,width=0.45\linewidth,clip=} &
\epsfig{file=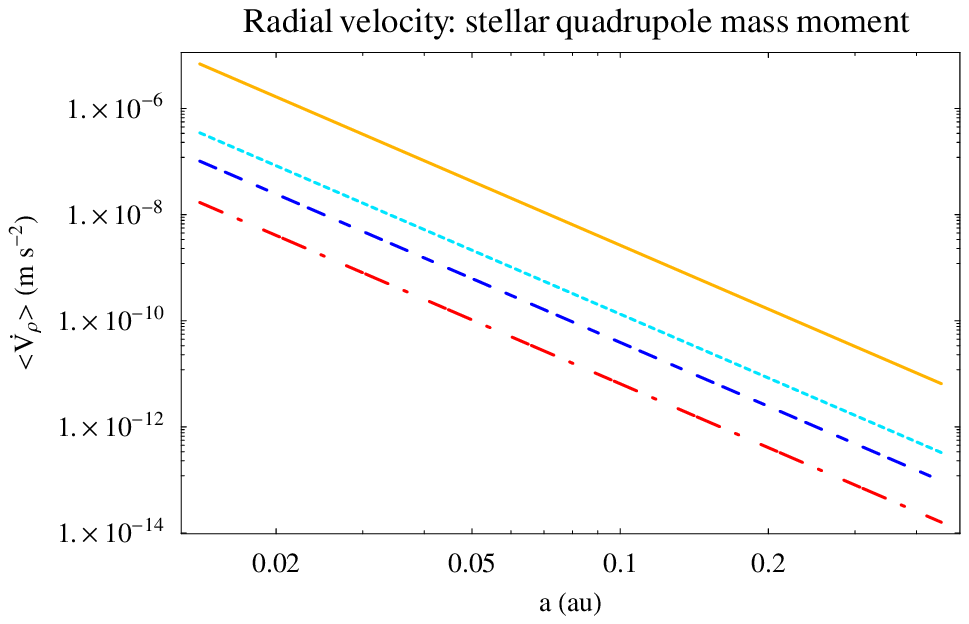,width=0.45\linewidth,clip=}
\end{tabular}
\caption{Maximum values of the long-term time variation $\left\langle\dot V_{\rho}\right\rangle$, in m s$^{-2}$, as a function of $a$ ($0.014\ {\rm au}\leq a\leq 0.449\ {\rm au}$) for different values of the eccentricity: $e=0.005$ (red dash-dotted line), $e=0.03$ (blue dashed line), $e=0.1$ (light blue dotted line), $e=0.6$ (yellow continuous line). For the stellar and planetary physical parameters, and for the inclination of the orbit to the plane of the sky and to the star/planet equators we  used the standard values of Table \ref{parametri}. We fixed the periastron at $\omega=90$ deg. }\lb{figurona2}
\end{figure*}
}
\subsection{Measurability of the long-term radial acceleration}
The present-day level of accuracy in measuring the radial velocity is about 2 m s$^{-1}$, so that over $\tau=10$ yr a net radial acceleration should be detectable at a $\approx 3\times 10^{-9}$ m s$^{-2}$ level. Such a guess is confirmed by the case of\footnote{\textrm{In addition to the planet HD 126614Ab ($a=2.35$ au, $e=0.41$, $P_{\rm b}=3.41$ yr), the star HD 126614A has also a faint M dwarf companion of $M_{\rm dw}\approx 0.3 M_{\odot}$ at a (sky-projected) distance of 33 au. Named HD 126614B, it was discovered with direct observations
using adaptive optics and the PHARO camera at Palomar Observatory \citep{Howa010}. An even more distant M dwarf, dubbed NLTT37349 and with $M_{\rm dw}\approx 0.2 M_{\odot}$, is also present in the system at a (sky-projected) distance of about $3\times 10^3$ au \citep{Chana}.}} HD 126614A, monitored for the past 10 yr, for which a secular change of the radial velocity \citep{Howa010} \eqi\left\langle\dot V^{\rm (meas)}_{\rho}\right\rangle=(5.13\pm 0.06)\times 10^{-7}\ {\rm m\ s}^{-2}\lb{superr}\eqf has been measured.
\textrm{Incidentally, let us note that for HD 126614A none of the \virg{internal} star-planet dynamical effects considered so far (tidal bulges, oblateness, general relativity) can explain\footnote{We assume $i=90$ deg.}  \rfr{superr}; indeed, the largest one, caused by the general relativistic gravitoelectric field, amounts to just $10^{-12}$ m s$^{-2}$. Concerning a distant third body X, it turns out that a Jupiter-sized planet at 2.2 au ($P_{\rm b}=3.0$ yr), assumed coplanar with HD 126614Ab in the equatorial plane of their hosting star, would cause a radial acceleration of the order of $5\times 10^{-7}$ m s$^{-2}$. Alternatively, a rocky Earth-like body at $0.32$ au ($P_{\rm b}=0.17$ yr) or a Neptune-type planet ($m_{\rm X}=14 m_{\oplus}$) at $0.75$ au ($P_{\rm b}=0.6$ yr) would induce the same effect.
It must be recalled that the quality of the fit of the 10 yr-long data record by \citet{Howa010} was not improved, at a statistically significant level, by a two-planet model for a variety of masses and distances for the second one.
A brown dwarf-like object ($m_{\rm X}=80 m_{\rm Jup}=7.6\times 10^{-2} M_{\odot}$) at 9 au ($P_{\rm b}=24.4$ yr) would also be a viable candidate.
According to \rfr{zomba}, neither NLTT37349 nor HD 126614B can accommodate \rfr{superr}. Indeed, the effect due to the closest M dwarf would be as large as $\lesssim 10^{-8}$ m s$^{-2}$, while that due to the farthest one is as small as $10^{-14}$ m s$^{-2}$. }

\section{The time variation of the temporal interval between primary and secondary transit}\lb{eclizza}
Another directly observable quantity in transiting exoplanets is the time elapsed $\Delta t_{\rm ecl}$ between the primary and the secondary eclipses. The primary eclipse occurs when the exoplanet starts to obscure the host star, while the beginning of the secondary eclipse is when the planet starts to be occulted by the star. For circular orbits, $\Delta t_{\rm ecl}$ is simply half the orbital period, while for eccentric orbits we have a more complicated expression involving $e$ and $\omega$ as well. It is \citep{Ste40,Jordan}
\eqi
\begin{array}{lll}
\Delta t_{\rm ecl}\doteq (t_{\rm 2ecl}-t_{\rm 1ecl})-\rp{P_{\rm b}}{2}  & = &\rp{P_{\rm b}}{\pi}\left[\rp{\sqrt{1-e^2}e\cos\omega}{1-\left(e\sin\omega\right)^2}+\right.\\ \\
&+&\left.\arctan\left(\rp{e\cos\omega}{\sqrt{1-e^2}}\right)\right],\lb{ecli}
\end{array}
\eqf up to terms proportional to $\cot^2 i$.
\subsection{Secular time variations of $\Delta t_{\rm ecl}$}\lb{sgalla}
Non-Keplerian orbital perturbations affect, in principle, $\Delta t_{\rm ecl}$ \textrm{by making it vary} orbit after orbit with a characteristic long-term pattern. It is possible to analytically work out exact expressions for  all the dynamical effects considered, apart from  the case of the distant third body X for which an approximate formula in $e$ is released.  \textrm{Such long-term harmonic effects} are
%\footnote{The expression for X has a singularity for $\Psi_{\star}=0,l_z\neq 0$.}
%
%
\begin{equation}
\begin{array}{lll}
\left\langle\left.\dert{\Delta t_{\rm ecl}}{t}\right|^{(J_2^{\star})}\right\rangle &=& -\left(\rp{R_{\star}}{a}\right)^2\rp{3eJ_2^{\star}  \left(3+5\cos 2\Psi_{\star}\right)\sin\omega}{2\sqrt{1-e^2}\left(1-e^2\sin^2\omega\right)^2},\lb{tuttoeclid}
\end{array}
\end{equation}
\begin{equation}
\begin{array}{lll}
\left\langle\left.\dert{\Delta t_{\rm ecl}}{t}\right|^{(j_2^{\rm p})}\right\rangle &=& -\left(\rp{r_{\rm p}}{a}\right)^2\rp{3ej_2^{\rm p}  \left(3+5\cos 2\psi_{\rm p}\right)\sin\omega}{2\sqrt{1-e^2}\left(1-e^2\sin^2\omega\right)^2},
\end{array}
\end{equation}
\begin{equation}
\begin{array}{lll}
\left\langle\left.\dert{\Delta t_{\rm ecl}}{t}\right|^{\rm(GE)}\right\rangle &=&-\left(\rp{\mathcal{R}_g}{a}\right)\rp{12 e\sqrt{1-e^2}\sin\omega}{\left(1-e^2\sin^2\omega\right)^2},
\end{array}
\end{equation}
\begin{equation}
\begin{array}{lll}
\left\langle\left.\dert{\Delta t_{\rm ecl}}{t}\right|^{\rm (GM)}\right\rangle &=&\left(\rp{GS_{\star}}{c^2 n a^3 }\right)\rp{24 e \cos\Psi_{\star} \sin\omega}{\left(1-e^2\sin^2\omega\right)^2},
\end{array}
\end{equation}
\begin{equation}
\begin{array}{lll}
\left\langle\left.\dert{\Delta t_{\rm ecl}}{t}\right|^{\rm (tid\ p)}\right\rangle &=&\\ \\
&=&\left(\rp{M_{\star}}{m_{\rm p}}\right)\left(\rp{r_{\rm p}}{a}\right)^5 \\ \\
& &\rp{15 e \left(-128 + 1152 e^2+6048 e^4+4200 e^6+441 e^8\right)k_{2{\rm p}} \sin\omega}{ 64(1-e^2)^{7/2} \left(1-e^2\sin^2\omega\right)^2},\lb{tuttoeclitidp}
\end{array}
\end{equation}
\begin{equation}
\begin{array}{lll}
\left\langle\left.\dert{\Delta t_{\rm ecl}}{t}\right|^{\rm (tid\ \star)}\right\rangle & = &\\ \\
&=&\left(\rp{m_{\rm p}}{M_{\star}}\right)\left(\rp{R_{\star}}{a}\right)^5 \\ \\
& &\rp{15 e \left(-128 + 1152 e^2+6048 e^4+4200 e^6+441 e^8\right)k_{2{\star}} \sin\omega}{ 64(1-e^2)^{7/2} \left(1-e^2\sin^2\omega\right)^2},
\lb{tuttoeclitids}
\end{array}
\end{equation}
\begin{equation}
\begin{array}{lll}
\left\langle\left.\dert{\Delta t_{\rm ecl}}{t}\right|^{\rm (X)}\right\rangle &=&\left(\rp{\mathcal{K}_{\rm X}}{n^2}\right)\rp{3e}{\left(1-e^2\right)^{3/2}\left(1-e^2\sin^2\omega\right)^3}\\ \\
& &\left[\sum_{k=1}^6 \left(l_i l_j\right)_k \mathfrak{S}_k\left(\Omega,\omega,\Psi_{\star}\right)\right]+\\ \\
&+&\mathcal{O}(e^2),
\end{array}\lb{tuttoeclix}
 \end{equation}
 where $\left(l_i l_j\right)_k, k=1\ldots 6$ are to be intended as the six products $l_x^2,l_x l_y, l_x l_z,l_y^2,l_y l_z,l_z^2$ of the components of $\bds{\hat{l}}_{\rm X}$, while  each of the six $\mathfrak{S}_k, k=1\ldots 6$ is a linear combination of trigonometric functions whose arguments are, in turn, linear combinations of $\Omega,\omega,\Psi_{\star}$.
 Note that \rfr{tuttoeclid}-\rfr{tuttoeclix} vanish in the limit $e\rightarrow 0$, as expected, while they do not vanish for $\Psi_{\star}=0$.

\textrm{It is interesting to note that the same considerations of Section \ref{sgulla} concerning the relative strengths of the tidally-induced effects on the radial velocity
hold also in this case. Indeed, according to \rfr{tuttoeclitidp}-\rfr{tuttoeclitids} the ratio of the planet-to-star tidal effects is equal just to \rfr{rakka}.
}
\subsection{Numerical evaluations}
For the typical star-planet scenario of Table \ref{parametri} we have
%with $a=0.04$ au, $e=0.07$ and $\Psi_{\star}=0$ we have
\begin{equation}
{\begin{array}{lll}
\left|\left\langle\left.\dert{\Delta t_{\rm ecl}}{t}\right|^{\rm(tid\ p)}\right\rangle\right| &\leq & 3\times 10^{-7},\\ \\
\left|\left\langle\left.\dert{\Delta t_{\rm ecl}}{t}\right|^{\rm(GE)}\right\rangle\right| &\leq & 2\times 10^{-7},\\ \\
\left|\left\langle\left.\dert{\Delta t_{\rm ecl}}{t}\right|^{(j_2^{\rm p})}\right\rangle\right| &\leq & 3\times 10^{-8},\\ \\
\left|\left\langle\left.\dert{\Delta t_{\rm ecl}}{t}\right|^{(J_2^{\star})}\right\rangle\right| &\leq & 2\times 10^{-9},\\ \\
\left|\left\langle\left.\dert{\Delta t_{\rm ecl}}{t}\right|^{\rm (tid\ \star)}\right\rangle\right| &\leq & 1\times 10^{-9},\\ \\
\left|\left\langle\left.\dert{\Delta t_{\rm ecl}}{t}\right|^{\rm (GM)}\right\rangle\right| &\leq & 4\times 10^{-11}, \\ \\
\end{array}}\lb{tuttoecli_numeri}
 \end{equation}
 The effect of a distant perturbing body X is comparable to, or even larger than \rfr{tuttoecli_numeri}. Indeed, for $m_{\rm X}=m_{\rm Jup}, r_{\rm X}=0.1-1$ au and $l_z=0$, i.e by assuming coplanarity with the perturbed planet, we have
\eqi \left|\left\langle\left.\dert{\Delta t_{\rm ecl}}{t}\right|^{\rm (Jup)}\right\rangle\right|\lesssim 1\times 10^{-5}-1\times 10^{-8}. \eqf
If we assume an Earth-sized perturbing body X at $0.1-1$ au we have
\eqi
\left|\left\langle\left.\dert{\Delta t_{\rm ecl}}{t}\right|^{\rm (Ear)}\right\rangle\right|\lesssim 4\times 10^{-8}-4\times 10^{-11}. \eqf

\textrm{
Figure \ref{figurona3} illustrates the magnitude of the upper bound of $\left\langle\dot \Delta t_{\rm ecl}\right\rangle$ for all the dynamical effects considered, with the exception of X, as a function of the semi-major axis for different values of the eccentricity.
\begin{figure*}
\centering
\begin{tabular}{cc}
\epsfig{file=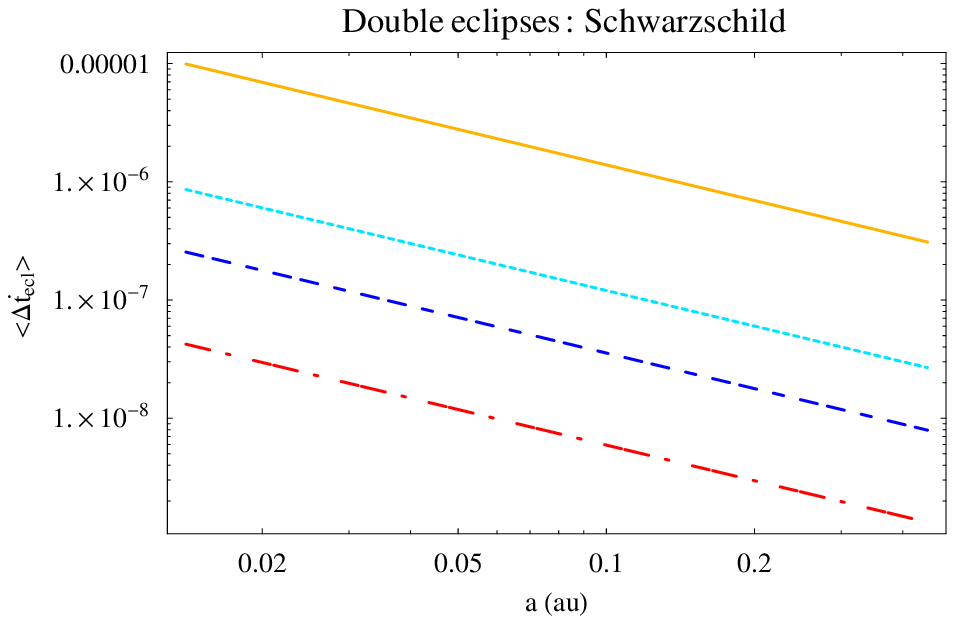,width=0.45\linewidth,clip=} &
\epsfig{file=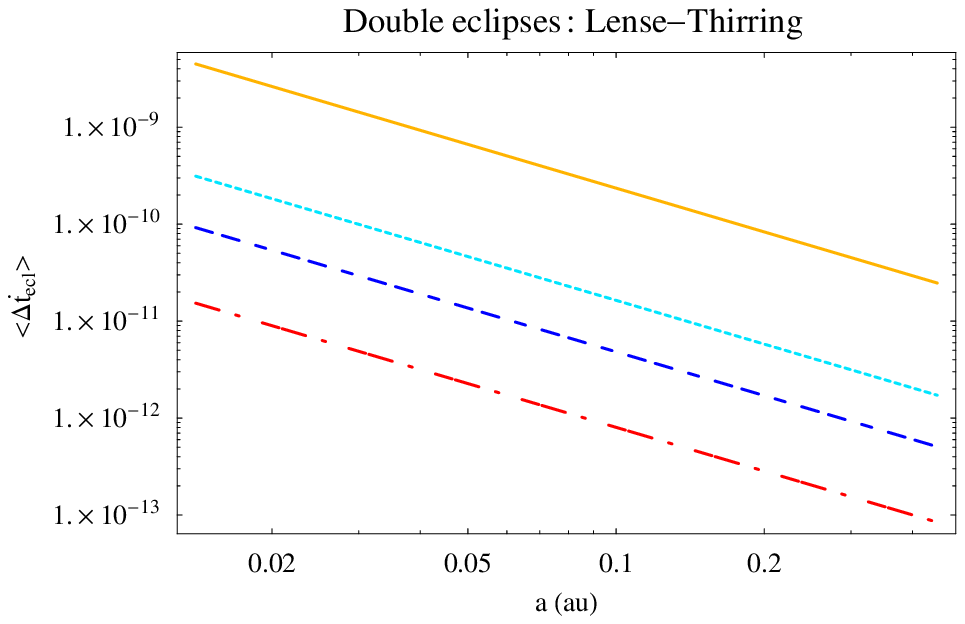,width=0.45\linewidth,clip=} \\
\epsfig{file=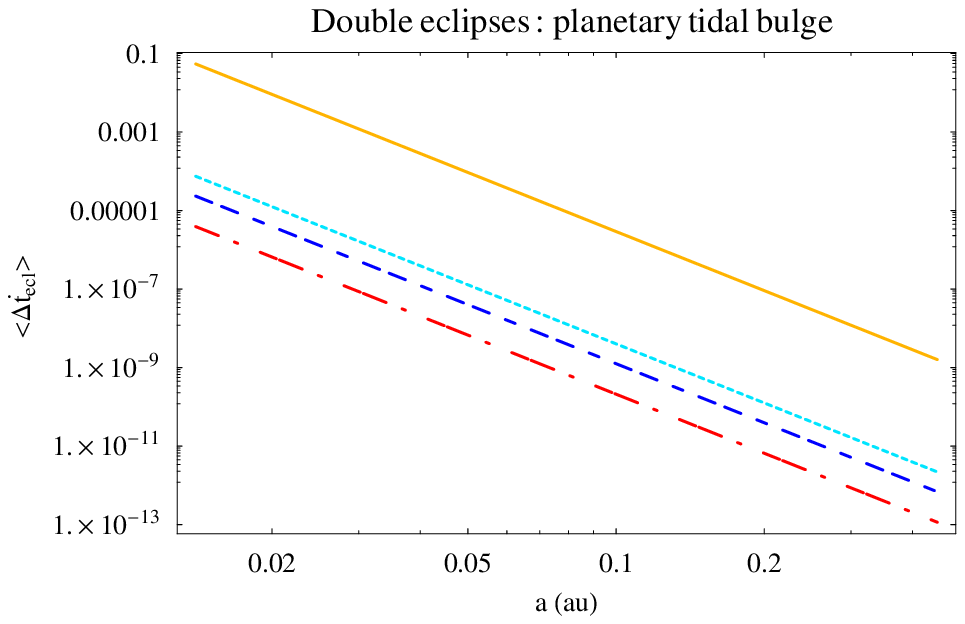,width=0.45\linewidth,clip=} &
\epsfig{file=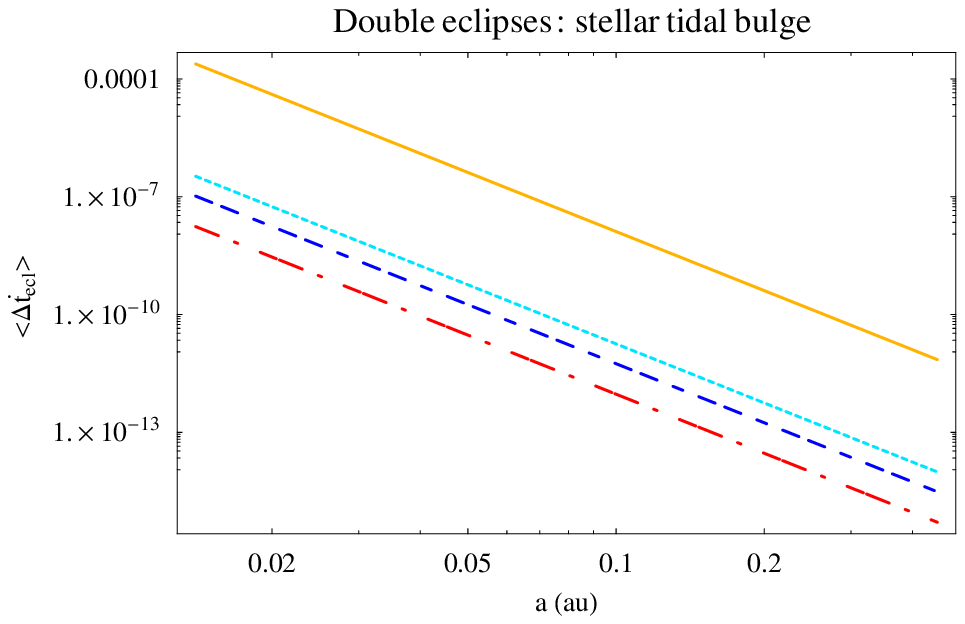,width=0.45\linewidth,clip=}\\
\epsfig{file=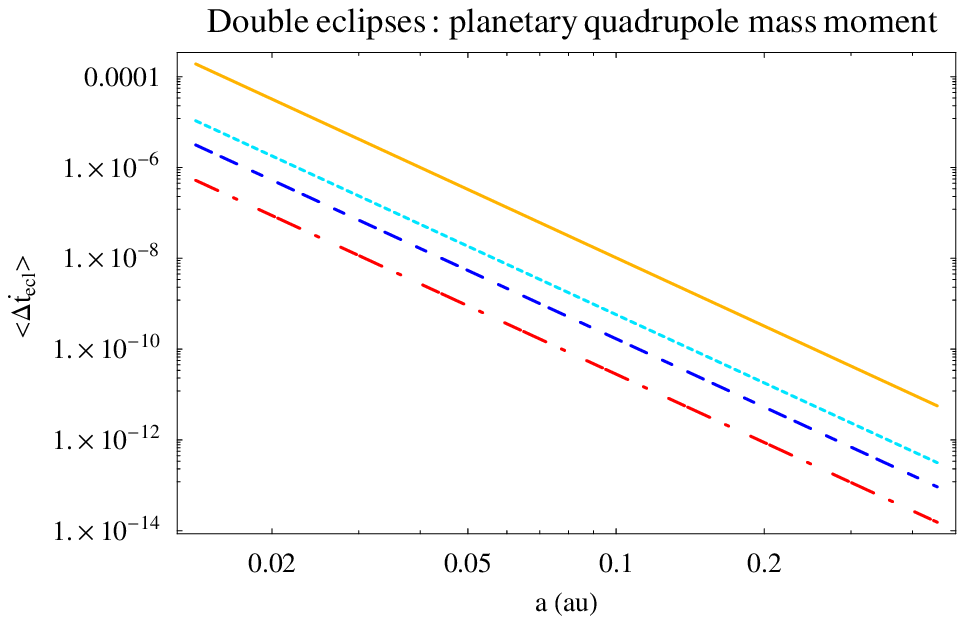,width=0.45\linewidth,clip=} &
\epsfig{file=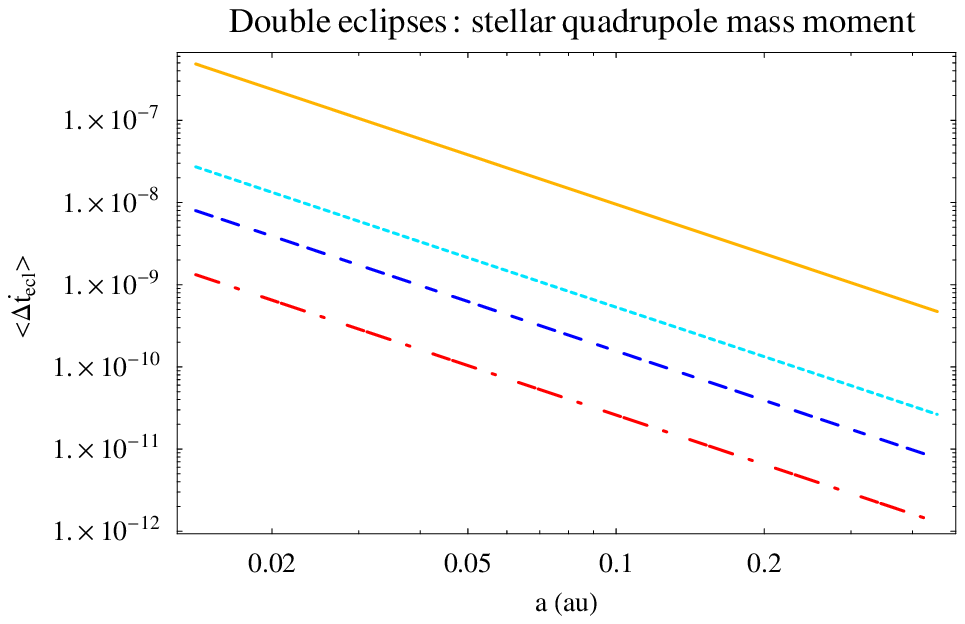,width=0.45\linewidth,clip=}
\end{tabular}
\caption{Maximum values of the long-term time variations  $\left\langle\dot\Delta t_{\rm ecl}\right\rangle$ as a function of $a$ ($0.014\ {\rm au}\leq a\leq 0.449\ {\rm au}$) for different values of the eccentricity: $e=0.005$ (red dash-dotted line), $e=0.03$ (blue dashed line), $e=0.1$ (light blue dotted line), $e=0.6$ (yellow continuous line). For the stellar and planetary physical parameters, and for the inclination of the orbit to the plane of the sky and to the star/planet equators we  used the standard values of Table \ref{parametri}. We fixed the periastron at $\omega=90$ deg. }\lb{figurona3}
\end{figure*}
}
\subsection{The measurability accuracy}
Concerning the measurability of $\Delta t_{\rm ecl}$, \citet{Jordan} point out that it would be dominated by the uncertainties in the time of the secondary eclipse $t_{\rm 2ecl}$ which is evaluated by them to be of the order of about $\sigma_{t_{\rm 2ecl}}\approx 80$ s.
Thus, it is reasonable to put
\eqi \sigma_{\Delta t_{\rm ecl} }\approx \rp{\sigma_{t_{\rm 2ecl}}}{\mathfrak{N_{\rm 2tr}}^{3/2}},\eqf where $\mathfrak{N}_{\rm 2tr}$ is the number of secondary transits observed. By posing $\tau=10$ yr and assuming $\mathfrak{N}_{\rm 2tr}=10^3$ it is possible to obtain
\eqi \sigma_{\dot\Delta t_{\rm ecl}}\approx 8\times 10^{-12}.\lb{incer}\eqf Thus, all the effects of \rfr{tuttoecli_numeri} would be measurable.

\section{The variation of the transit period in elliptic orbits}\lb{titivu}
\textrm{Another important directly measurable quantity in transiting exoplanets is the temporal interval $P_{\rm tr}$ between successive primary transits \citep{Miralda,Agol,Nes1,Nes2,Veras,Payne}.
To compute the variation of the transit period $P_{\rm tr}$ in elliptic orbits we will make use of the so-called guiding center approximation \citep{Murray}. In it, the Keplerian motion of a test particle p moving along an elliptic orbit with focus F is described in a suitable rotating reference frame $\{x^{'} y^{'} \}$. Referring to Figure 2.8 of \citet{Murray} for the orientation of the axes, the guiding center G is the origin of such a frame. It moves with  angular speed equal to the mean motion $n$ along a circle centered in F ad having radius equal to the semi-major axis $a$. By orienting the $y^{'}$ axis tangentially
the guiding center's circle towards the motion of the test particle and the $x^{'}$ axis perpendicularly to it from F to p, it turns out \citep{Murray}
 \begin{equation}
{\begin{array}{lll}
x^{'}&=&r\cos(f-\mathcal{M})-a,\\ \\
y^{'}&=&r\sin(f-\mathcal{M}).
\end{array}}\lb{epicyclic}
 \end{equation}
 Concerning the argument of the harmonic functions in \rfr{epicyclic}, the exact relation between $f$ and $\mathcal{M}$ can be derived from \rfr{yta}. It is \citep{Capde}
 \eqi \mathcal{M}=2\arctan\left[\sqrt{\rp{1-e}{1+e}}\tan\left(\rp{f}{2}\right)\right]-\rp{e\sqrt{1-e^2}\sin f}{(1+e\cos f)}.\lb{maronna}\eqf
 A useful approximation of \rfr{maronna} is \citep{Roy}
 \eqi \mathcal{M}= f -2e\sin f+\rp{3}{4}e^2\sin 2f+\mathcal{O}(e^3).\lb{madonnina}\eqf
 As far as the transit period $P_{\rm tr}$ is concerned, from the choice of the axes in the rotating frame it turns out that the component relevant to such a phenomenon is $y^{'}$.
Thus, the departure of $P_{\rm tr}$ from the Keplerian orbital period $P_{\rm b}\doteq 2\pi/n$ is due to the overall change of $y^{'}$ over one orbital revolution. Since the linear speed of the guiding center is $v_{\rm G}=na$,
\eqi P_{\rm tr}=P_{\rm b}+\rp{\Delta y^{'}}{v_{\rm G}}=\rp{2\pi}{n}\left(1+\lambda_{\rm tr}\right),\lb{transito}\eqf
with
\eqi \lambda_{\rm tr}\doteq\rp{\Delta y^{'}}{2\pi a},\eqf
in which $\Delta y^{'}$ is the variation of $y^{'}$ per orbit, i.e. integrated over one orbital period.
%Taking into account possible variations of $a$ an $e$ as well due to orbital perturbations,  we have, for a generic position $f$,
%\eqi\Delta y^{'}(f)=\int_0^f %dy^{'}(\tilde{f})=\int_0^f\left[\derp{y^{'}}{a}\dert{a}{t}+\derp{y^{'}}{e}\dert{e}{t}+\derp{y^{'}}{\tilde{f}}\dert{\tilde{f}}{t}\right]\left(\dert{t}{\tilde{f}}\right)d\tilde{f},\lb{cacau}\eqf
%where $dt/df$ is given by \rfr{keplo}.
%Quite generally,
%\eqi dy^{'}=dy^{'}_{\rm Kep}+dy^{'}_{\rm pert}= dy^{'}_{\rm Kep}+\left(\mathcal{C}A_R +\mathcal{D}A_T\right)df.\lb{pizza}\eqf
%The exact analytical expressions of $dy^{'}_{\rm Kep},\mathcal{C},\mathcal{D}$ are quite cumbersome, and we will not show them here.
%A first check of the correctness of our approach consists of testing if the Keplerian variation of $y^{'}$ per orbit does actually vanish, as expected.
%A numerical integration making use of the exact result of \rfr{maronna} shows that
%\eqi \int_0^{2\pi}dy^{'}_{\rm Kep}=0.\eqf
By using the first term of order $\mathcal{O}(e)$ in the approximation of \rfr{madonnina} and\footnote{As a consequence, p moves about G in the opposite sense with respect to G about F on a 2:1 ellipse of period $2\pi/n$ \citep{Murray}.}
\eqi y^{'}\approx 2 a e \sin f,\lb{yzb}\eqf
it is possible to
%obtain
%
%
%\begin{equation}
%{\begin{array}{lll}
%\mathcal{C}&=& \rp{2(1-e^4)\sin^2 f}{n^2\left(1+e\cos f\right)^2},\\ \\
%
%\mathcal{D}&=&\rp{(1-e^2) \left[e\left(7-e^2\right)+\left(1+e^2\right)\left(4\cos f+e\cos 2f\right)\right]\sin f}{n^2\left(1+e\cos f\right)^3}.
%\end{array}}\lb{periocoff}
% \end{equation}
%Such relations will allow to
explicitly compute the effects of several dynamical perturbations on $P_{\rm tr}$.
The averaged time variation of it can  straightforwardly be computed from \rfr{ypa} for $Y\rightarrow y^{'}/na$, with $y^{'}$ conveniently given by \rfr{yzb}.
A common feature of the results that we are going to show below is that non-zero long-term variations of $P_{\rm tr}$ occur for all the dynamical effects considered; they do not vanish in the limit $e\rightarrow 0$.
\subsection{The effect of the stellar oblateness}
A straightforward calculation yields the following non-zero long-term harmonic effect.
\eqi
\begin{array}{lll}
\left\langle \left.\dert{P_{\rm tr}}{t}\right|^{(J_2^{\star})}\right\rangle &=& -\rp{3 J_2^{\star}}{64(1-e^2)^3}\left(\rp{R_{\star}}{a}\right)^2\times \\ \\
&\times &\left[-(2+e^2)(-8 + 5e^2)(1+3\cos 2\Psi_{\star})-\right.\\ \\
&-&\left. e^2\left(8+\rp{77}{2}e^2\right)\sin^2\Psi_{\star}\cos2\omega\right].\lb{refj2}
\end{array}
\eqf
Note that \rfr{refj2} vanish neither for equatorial ($\Psi_{\star}=0$) nor for polar ($\Psi_{\star}=90$ deg) orbits.
\subsection{The tidal bulges}
It is found that the tidal bulges cause non-vanishing long-term secular\footnote{Indeed, $a$ and $e$ do not undergo long-term time variations.} variations of $P_{\rm tr}$. They are
\eqi
\begin{array}{lll}
\left\langle \left.\dert{P_{\rm tr}}{t}\right|^{(\rm tid\ p)}\right\rangle &=&\left(\rp{M_{\star}}{m_{\rm p}}\right)\left(\rp{r_{\rm p}}{a}\right)^5 \rp{3\left(-128-464e^2+320 e^4 +195 e^6\right)k_{\rm 2p}}{64(1-e^2)^6},\lb{reftidp}
\end{array}
\eqf
\eqi
\begin{array}{lll}
\left\langle \left.\dert{P_{\rm tr}}{t}\right|^{(\rm tid\ \star)}\right\rangle &=&\left(\rp{m_{\rm p}}{M_{\star}}\right)\left(\rp{R_{\star}}{a}\right)^5 \rp{3\left(-128-464e^2+320 e^4 +195 e^6\right)k_{2\star}}{64(1-e^2)^6}.\lb{reftids}
\end{array}
\eqf
Note that, also in this case, the ratio of the planet-to-star tidal effects is equal just to \rfr{rakka}, as in Section \ref{sgulla} for $V_{\rho}$ and Section \ref{sgalla} for $\Delta t_{\rm ecl}$.
\subsection{The effects of general relativity}
The general relativistic gravitoelectric term induces the following long-term secular variation of $P_{\rm tr}$.
\eqi
\left\langle \left.\dert{P_{\rm tr}}{t}\right|^{(\rm GE)}\right\rangle = \left(\rp{\mathcal{R}_g}{a}\right)\rp{(24+33e^2-7e^4)}{4(1-e^2)^2}.\lb{refge}
\eqf
It is interesting to note that \rfr{refge} is quite different from eq.(17) by \citet{Jordan} yielding an instantaneous formula proportional to $e$.
Also the gravitomagnetic field causes a net non-vanishing harmonic effect, which is
\eqi
\left\langle \left.\dert{P_{\rm tr}}{t}\right|^{(\rm GM)}\right\rangle = \left(\rp{GS_{\star}}{c^2 n a^3}\right)\rp{2\left(2-\rp{e^2}{4}\right)\cos\Psi_{\star}}{(1-e^2)^{3/2}}.\lb{refgm}
\eqf
It vanishes for polar orbits.
\subsection{The third body X}
As expected, in this case the calculations are much more cumbersome. To effectively perform them, the further approximation
\eqi (1+e\cos f)^{-k}\approx 1-ke\cos f,\ k=2,3,4\eqf
is adopted.
We have
\eqi
\left\langle \left.\dert{P_{\rm tr}}{t}\right|^{(\rm X)}\right\rangle =\left(\rp{\mathcal{K}_{\rm X}}{ n^2}\right)\rp{3\left(1-e^2\right)^2\mathfrak{Y}(\bds{\hat{l}}_{\rm X},\Omega,\omega,\Psi_{\star})}{16},\lb{refX}
\eqf
where $\mathfrak{Y}$ is a complicated function of the angular orbital parameters of the planet and of the position of X. As in the previous cases, it is not worth explicitly showing it.
It is found that the long-term harmonic signal of \rfr{refX} does not vanish for any particular orbital configurations of both the perturbed planet p and the distant perturber X. For example, if both lie in the equatorial plane of the hosting star, \rfr{refX} is not zero.
\subsection{Numerical evaluations}\lb{numeracci}
The standard scenario of Table \ref{parametri} yields
\begin{equation}
{\begin{array}{lll}
\left|\left\langle\left.\dert{P_{\rm tr}}{t}\right|^{\rm(GE)}\right\rangle\right| &= & 1\times 10^{-6},\\ \\
\left|\left\langle\left.\dert{P_{\rm tr}}{t}\right|^{\rm(tid\ p)}\right\rangle\right| &= & 8\times 10^{-7},\\ \\
\left|\left\langle\left.\dert{P_{\rm tr}}{t}\right|^{(j_2^{\rm p})}\right\rangle\right| &= & 1\times 10^{-7},\\ \\
\left|\left\langle\left.\dert{P_{\rm tr}}{t}\right|^{(J_2^{\star})}\right\rangle\right| &= & 8\times 10^{-9},\\ \\
\left|\left\langle\left.\dert{P_{\rm tr}}{t}\right|^{\rm (tid\ \star)}\right\rangle\right| &= & 4\times 10^{-9},\\ \\
\left|\left\langle\left.\dert{P_{\rm tr}}{t}\right|^{\rm (GM)}\right\rangle\right| &= & 1\times 10^{-10}, \\ \\
\end{array}}\lb{tuttottv_numeri}
 \end{equation}
The effect of a Jupiter-sized third body X at $r_{\rm X}=0.1-1$ au is of the order of $10^{-5}-10^{-8}$. An Earth-type perturber X at the same distances
would yield variations of $P_{\rm tr}$ of about $10^{-8}-10^{-11}$.
Figure \ref{figurona4} depicts the magnitude of the upper bound of $\left\langle\dot P_{\rm tr}\right\rangle$ for all the dynamical effects considered, with the exception of X, as a function of the semi-major axis for different values of the eccentricity.
\begin{figure*}
\centering
\begin{tabular}{cc}
\epsfig{file=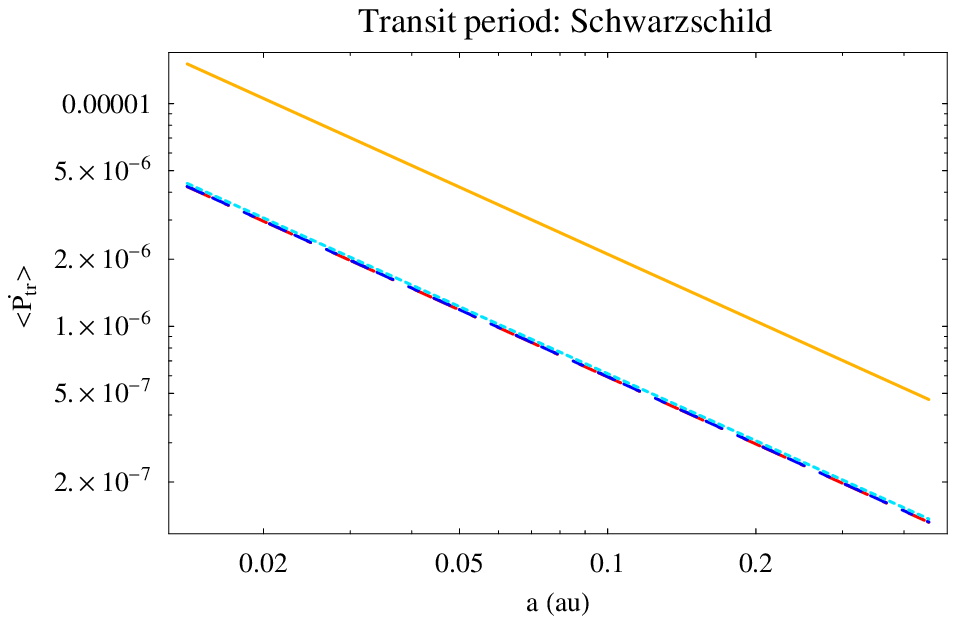,width=0.45\linewidth,clip=} &
\epsfig{file=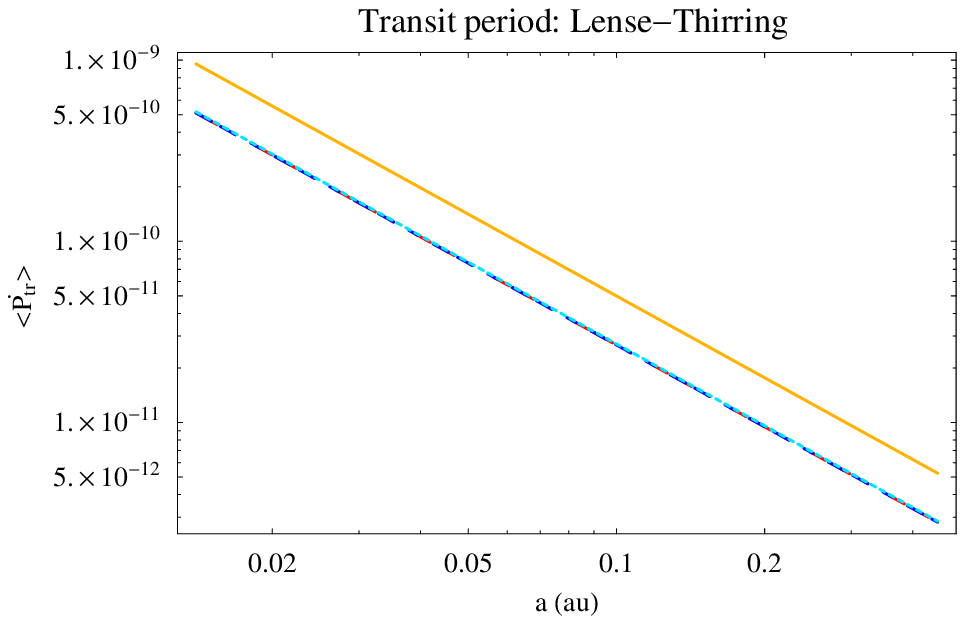,width=0.45\linewidth,clip=} \\
\epsfig{file=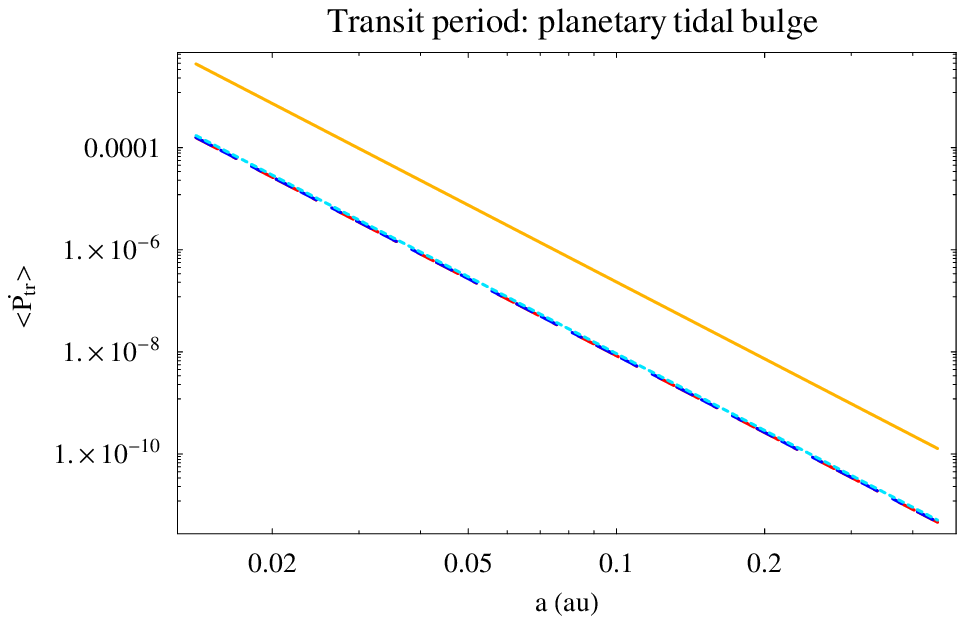,width=0.45\linewidth,clip=} &
\epsfig{file=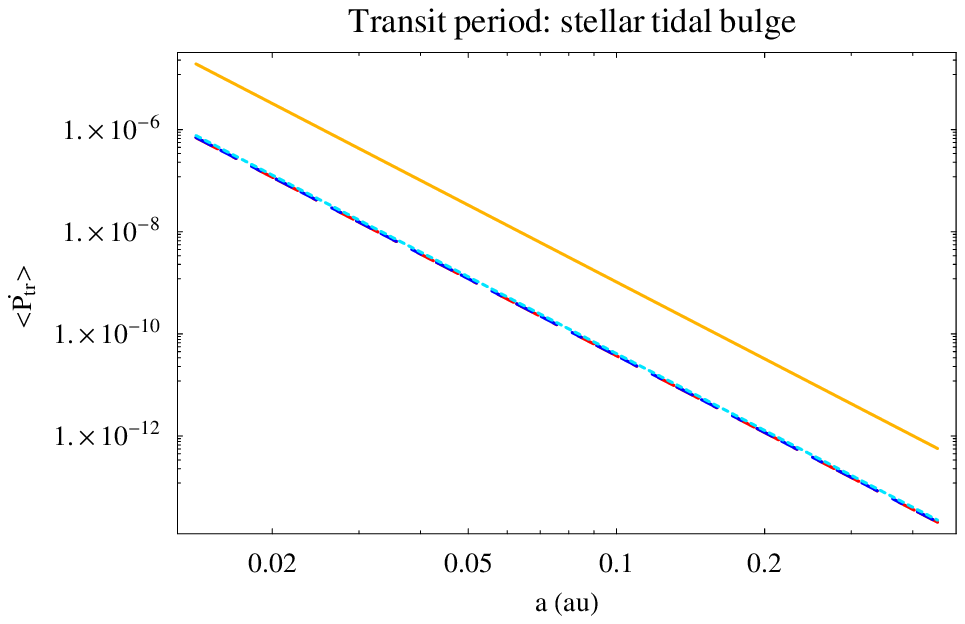,width=0.45\linewidth,clip=}\\
\epsfig{file=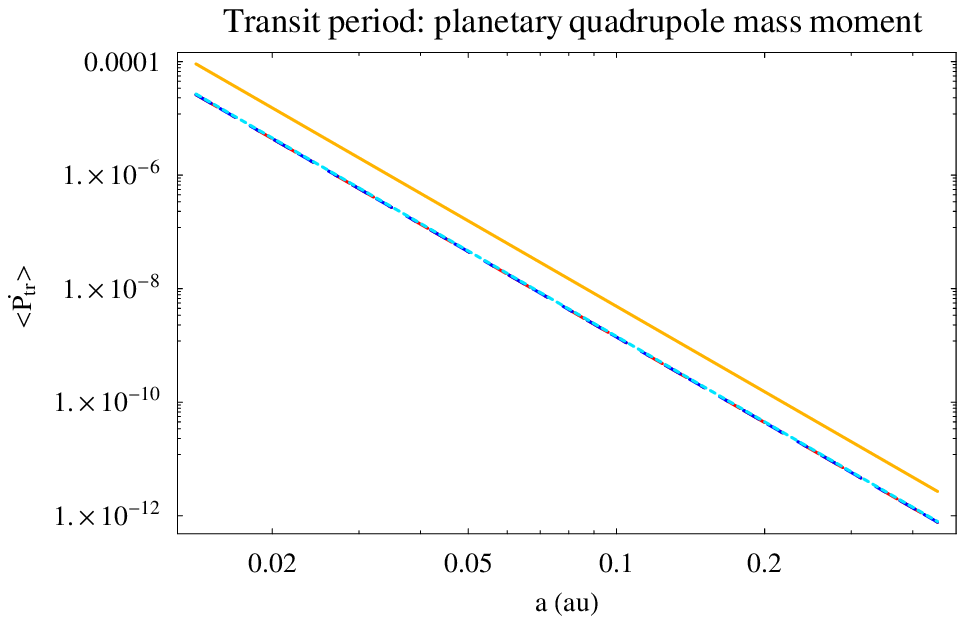,width=0.45\linewidth,clip=} &
\epsfig{file=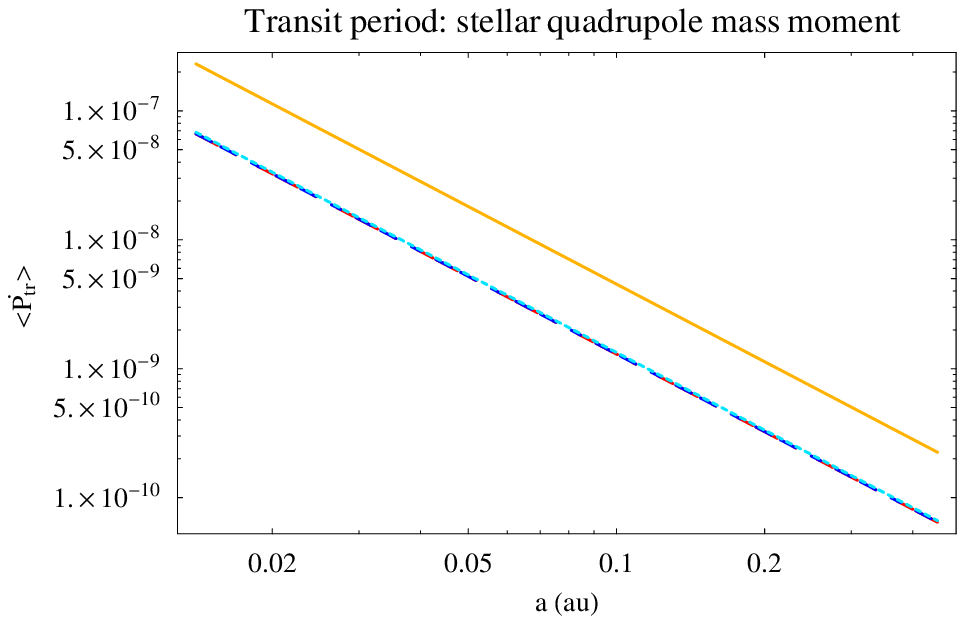,width=0.45\linewidth,clip=}
\end{tabular}
\caption{Maximum values of the long-term time variations $\left\langle\dot P_{\rm tr}\right\rangle$ as a function of $a$ ($0.014\ {\rm au}\leq a\leq 0.449\ {\rm au}$) for different values of the eccentricity: $e=0.005$ (red dash-dotted line), $e=0.03$ (blue dashed line), $e=0.1$ (light blue dotted line), $e=0.6$ (yellow continuous line). For the stellar and planetary physical parameters, and for the inclination of the orbit to the plane of the sky and to the star/planet equators we  used the standard values of Table \ref{parametri}. We fixed the periastron at $\omega=90$ deg. }\lb{figurona4}
\end{figure*}
\subsection{Measurement accuracy}
Concerning the accuracy with which  time variations of the transit period could be detected, \citet{Miralda} yields $\sigma_{\dot P_{\rm tr}}\approx 10^{-13}$ over $\tau=10$ yr. He assumes about $\mathfrak{N}_{\rm 1tr}=1000$ primary transits and an accuracy in measuring each primary transit of $\sigma_{\rm 1ecl}\approx 1$ s. A similar evaluation is given by \citet{Jordan}. Note that their formula of eq.(17) yields a value for the general relativistic gravitoelectric effect which is about 6 orders of magnitude smaller than ours in \rfr{tuttottv_numeri}; the pessimistic conclusions by \citet{Jordan}  are actually based just on such a small figure.
Finally, let us note that \citet{Macie} have recently detected transit timing variations in WASP-3b ($a=0.03$ au, $e\approx 0$, $P_{\rm b}=1.8$ d) over $\tau=2$ yr (2009-2010), but they exhibit a harmonic pattern with a periodicity of about 124 d and semi-amplitude of $1.4\times 10^{-3}$ d, with no discernable secular trends. In fact, such kind of cumulatively growing patterns, which may be caused by the tidal bulges and the general relativistic gravitoelectric field according to \rfr{reftidp}-\rfr{reftidp} and \rfr{refge}, could not have been detected since the accuracy with which $P_{\rm tr}$ has been measured by \citet{Macie} is of the order of $2.4 P_{\rm b}=4 \times 10^5$ s. Over $\tau=2$ yr this naively translates into a potential accuracy in detecting $\dot P_{\rm tr}$ of about $6\times 10^{-3}$. Looking at \rfr{refX} for X, it results that long-period harmonic effects are, in principle, possible in view of the interplay between the node and the periastron of WASP-3b entering $\mathfrak{Y}$; recall that they slowly change because of the oblateness, tidal bulges, general relativity and a X themselves.
However, the figures for $\left\langle\dot P_{\rm tr}\right\rangle $ in Section \ref{numeracci}, valid for plausible values of $\mathcal{K}_{\rm X}$,
 do not allow to obtain magnitudes as large as\footnote{They can be obtained by multiplying $\left\langle\dot P_{\rm tr}\right\rangle $ by $P_{\rm b}$.} $10^{-3}$ d.}
\section{Summary and conclusions}\lb{concluzza}
We looked at the impact of several classical and general relativistic dynamical perturbations on directly observable quantities in transiting exoplanets.
We considered the centrifugal oblateness of both the hosting star and the  planet due to their rotations, the tidal bulges mutually raised by both the star and the planet on each other, and a distant third body X as Newtonian effects. Concerning  general relativity,  we took into account both the  \textrm{so called gravitoelectric}, Schwarzschild and the  \textrm{gravitomagnetic}, Lense-Thirring perturbations.
 The observables considered are
 the transit duration $\Delta t_d$, the radial velocity $V_{\rho}$, the time interval $\Delta t_{\rm ecl}$ elapsed between primary and secondary eclipses, \textrm{and the time span $P_{\rm tr}$ between successive primary transits.}

 We analytically worked out, in an uniform and straightforward way, the long-term, i.e. averaged over one orbital revolution, temporal variations of such quantities caused by the aforementioned non-Keplerian features of motion. We did not restrict ourselves to circular orbits.

 For all the \textrm{dynamical} effects considered we released numerical evaluations of their magnitudes by adopting a typical exoplanetary scenario involving a Jupiter-sized planet closely ($a=0.04$ au) revolving around a main sequence Sun-like star along a moderately eccentric  orbit ($e=0.07$) lying in the orbital plane of the parent star ($\Psi_{\star}=0$). The rotation of the planet has been assumed synchronized with its orbital frequency because of tidal effects, and an edge-on configuration ($i=90$ deg) has been assumed.
 For each observable considered, we gave an order-of-magnitude evaluation of the accuracy with which it could be measured over an observational time span $\tau=10$ yr. \textrm{In view of the wide distribution of orbital configurations which is expected to characterize the numerous planets which will be discovered by the ongoing space-based Kepler mission, we also performed graphical investigations of the dependence of the effects studied on the semi-major axes $a$ and the eccentricities $e$.}

 No net changes $\left\langle\dot\Delta t_{d}\right\rangle$ in the duration transit occur for both the tidal bulges and the Schwarzschild perturbations. Instead, the oblateness of the bodies, a distant planet X and the Lense-Thirring effect induce  \textrm{long-term, harmonic variations of the} transit duration which do not vanish \textrm{even} for circular orbits. For exactly edge-on orbits ($i=90$ deg), $\left\langle\dot\Delta t_{d}\right\rangle=0$ in all cases. The same occur, independently of the inclination $i$ of the orbital plane to the line of sight, for equatorial orbits ($\Psi_{\star}=0$), apart from X if it is not coplanar with the perturbed body. If both the planet and its perturber X lie in the same plane coinciding with the equatorial plane of the star, then $\left\langle\dot\Delta t_{d}\right\rangle=0$. By assuming small departures from the equatorial ($\Psi_{\star}=15$ deg) and edge-on ($i=87$ deg) orbital configuration, the magnitude of the upper limits of the non-zero oblateness and gravitomagnetic effects is of the order of $10^{-8}-10^{-12}$. A jovian perturbing body X at $0.1-1$ au would cause a rate of $\approx 10^{-5}-10^{-8}$, while for an Earth-size X at the same distances we have $\left\langle\dot\Delta t_{d}\right\rangle\approx 10^{-7}-10^{-10}$. The expected accuracy  in measuring $\left\langle\dot\Delta t_d\right\rangle$ over $\tau=10$ yr is about $\sigma_{\dot\Delta t_d}\approx 10^{-9}-10^{-8}$.

 The radial velocity $V_{\rho}$ undergoes non-vanishing long-term, \textrm{harmonic variations}  caused by all the perturbations considered only for eccentric orbits; the resulting \textrm{amplitudes} are proportional to $e$, and vanish neither for equatorial nor edge-on orbits. The largest effects, of the order of $10^{-7}$ m s$^{-2}$, are due to the tidal bulge raised on the planet and the general relativistic Schwarzschild term. The centrifugal oblateness of the planet  causes an acceleration of the order of $10^{-8}$ m s$^{-2}$, while the changes due to the centrifugal oblateness and the tidal distortion of the star are of the order of $10^{-9}$ m s$^{-2}$. The Lense-Thirring effect amounts to $\approx 10^{-11}$ m s$^{-2}$. The perturbation of a distant Jupiter-sized  body X at $0.1-1$ au is $\left\langle\dot V_{\rho}\right\rangle\approx 10^{-5}-10^{-8}$ m s$^{-2}$, while a rocky planet with the mass of the Earth located at the same distances would cause $\left\langle\dot V_{\rho}\right\rangle\approx 10^{-8}-10^{-11}$ m s$^{-2}$. The accuracy with which it is possible to measure secular trends in the radial velocity is $\sigma_{\dot V_{\rho}}\approx 10^{-9}$ m s$^{-2}$ over $\tau=10$ yr.

Concerning the time interval $\Delta t_{\rm ecl}$ between primary and secondary eclipses, it \textrm{experiences long-term, harmonic variations}  only for eccentric orbits. Its \textrm{amplitudes} are proportional to $e$, and are non-vanishing for all the perturbations considered.
Also in this case, the largest effects, of the order of $10^{-7}$, are due to the tidal bulge raised on the planet and the general relativistic  \textrm{gravitoelectric field}. The centrifugal oblateness of the planet  causes an \textrm{effect} of the order of $10^{-8}$, while the variations induced by the centrifugal oblateness and the tidal distortion of the star are of the order of $10^{-9}$. \textrm{The effect due to the general relativistic gravitomagnetic field}  is as large as about $10^{-11}$. The perturbation of a jovian-type  body X at $0.1-1$ au is $\left\langle\dot\Delta t_{\rm ecl}\right\rangle\approx 10^{-5}-10^{-8}$, while a terrestrial planet at the same distances would induce $\left\langle\dot\Delta t_{\rm ecl}\right\rangle\approx 10^{-8}-10^{-11}$.
The expected accuracy in measuring $\left\langle\dot\Delta t_{\rm ecl}\right\rangle$ over $\tau=10$ yr is about $\sigma_{\dot\Delta t_{\rm ecl}}\approx 10^{-12}$.

\textrm{All the dynamical effects considered causes long-term variations of the transit period $P_{\rm tr}$. They do not vanish for circular orbits. While the signals due to the stellar and planetary oblateness, the gravitomagnetic  field and a third body X are harmonic, those induced by the tidal bulges and the gravitoelectric field are secular rates. The largest effect, of the order of $10^{-6}$, is the trend due to general relativity (Schwarzschild).
The rate due to the planetary tidal bulge is $8\times 10^{-7}$, while that due to the star's tidal distortion is 2 orders of magnitude smaller. The amplitudes of the signals caused by the planetary and the stellar oblateness are $1\times 10^{-7}$ and $8\times 10^{-9}$, respectively. The gravitomagnetic  effect has an amplitude of the order of $10^{-10}$.
The action of a distant Jupiter-like  planet X at $0.1-1$ au is $\left\langle\dot P_{\rm tr}\right\rangle\approx 10^{-5}-10^{-8}$, while an Earth-type planet at the same distances would cause $\left\langle\dot P_{\rm tr}\right\rangle\approx 10^{-8}-10^{-11}$.
It should be possible to measure long-term changes in $P_{\rm tr}$ with an accuracy of about $10^{-13}$ over $\tau=10$ yr.
}

Our results may be useful in focussing the research on some specific exoplanets or classes of exoplanets, particularly tailored to measure a given dynamical effect of interest. Moreover, they should allow to better evaluate the systematic uncertainty in the determination of a given stellar/planetary physical parameter induced by the other perturbations which are to be considered as sources of bias.
Moreover, our approach can, in principle,  be extended also to other scenarios and dynamical features of motion involving, e.g., natural and artificial bodies in our solar system in order to suitably design new tests of general relativity and modified models of gravity. The latter issues\textrm{, along with a more detailed quantitative analysis of different, specific exoplanetary scenarios,} may be the subject of further researches.

\textrm{
\section*{Acknowledgments}
I gratefully thank the referee for valuable suggestions and remarks.
}
%-----------------------------------------

\end{document}